\pdfoutput=1
\documentclass[12pt,a4paper]{article}
\usepackage{jheppub}
\usepackage[T1]{fontenc} 
\usepackage{graphicx}
\usepackage{epsfig}
\usepackage{rotating}
\usepackage{amssymb}
\usepackage{dsfont}
\usepackage{psfrag}
\usepackage{amsmath,euscript,array,mathrsfs,eufrak}
\usepackage{bbold}
\usepackage{epsf}
\usepackage{slashed}
\usepackage{cancel}
\usepackage{float}
\usepackage{caption,subcaption,wrapfig}
\usepackage[colorlinks=true,linktocpage=true,linkcolor=blue,citecolor=blue]{hyperref}
\usepackage{tikz}
\usetikzlibrary{decorations.markings,decorations.pathmorphing}

\newcommand{\be}{\begin{equation}}
\newcommand{\ee}{\end{equation}}
\newcommand{\beqs}{\begin{eqnarray}}
\newcommand{\eeqs}{\end{eqnarray}}

\newcommand{\pat}{\partial}
\newcommand{\TCP}{T_{\text{\tiny{CP}}}}
\newcommand{\alphaEE}{\alpha_{\text{\tiny{EE}}}}

\newcommand{\dd}{\mathrm{d}}

\newcommand{\Lone}{\Lambda_1}
\newcommand{\Ltwo}{\Lambda_2}

\newcommand{\OO}{\mathcal{O}}

\newcommand{\NN}{\mathcal{N}}

\newcommand{\B}{\mathds{B}}

\newcommand{\R}{\mathds{R}}
\newcommand{\CP}{\mathds{C}\mathds{P}}
\newcommand{\parent}[1]{\left(#1\right)}

\newcommand{\ls}{\ell_s}
\newcommand{\gs}{g_s}

\newcommand{\btriple}{b_0^{\text{\tiny triple}}}
\newcommand{\bcritical}{b_0^{\text{\tiny critical}}}

\newcommand{\SEE}{\mathsf{S}}

\makeatletter
\def\@hex@@Hex#1%
{\if a#1A\else \if b#1B\else \if c#1C\else \if d#1D\else
	\if e#1E\else \if f#1F\else #1\fi\fi\fi\fi\fi\fi \@hex@Hex}
\makeatother

\definecolor{c1}{HTML}{bd4008}
\definecolor{c2}{HTML}{f78026}
\definecolor{c3}{HTML}{fdbe11}
\definecolor{c4}{HTML}{44c721}
\definecolor{c5}{HTML}{1cad85}
\definecolor{c6}{HTML}{2792b6}
\definecolor{c7}{HTML}{025394}
\definecolor{c8}{HTML}{1d3585}
\definecolor{c9}{rgb}{.5,0,.5}

\graphicspath{{figures/}}
\numberwithin{equation}{section}

\title{Limitations of entanglement entropy \\ in detecting thermal phase transitions}

\author[1,2]{Niko Jokela,}
\author[1,2]{Helime Ruotsalainen,} 
\author[3,4]{and Javier G. Subils}

\affiliation[1]{Department of Physics, P.O. Box 64, FI-00014 University of Helsinki, Finland}
\affiliation[2]{Helsinki Institute of Physics P.O. Box 64, FI-00014 University of Helsinki, Finland}
\affiliation[3]{ Nordita, Stockholm University and KTH Royal Institute of Technology,\\
    Hannes Alfv\'ens v\"ag 12, SE-106 91 Stockholm, Sweden}
\affiliation[4]{ Institute for Theoretical Physics, Utrecht University, 3584 CC Utrecht, The Netherlands}

\emailAdd{niko.jokela@helsinki.fi}
\emailAdd{helime.ruotsalainen@helsinki.fi}
\emailAdd{javier.subils@su.se}

\date{\today}

\abstract{We explore the efficacy of entanglement entropy as a tool for detecting thermal phase transitions in a family of gauge theories described holographically. The rich phase diagram of these theories encompasses first and second-order phase transitions, as well as a critical and a triple point. While entanglement measures demonstrate some success in probing transitions between plasma phases, they prove inadequate when applied to phase transitions leading to gapped phases. Nonetheless, entanglement measures excel in accurately determining the critical exponent associated with the observed phase transitions, providing valuable insight into the critical behavior of these systems.}

\preprint{$\begin{array}{rr}\text{HIP-2023-15/TH}\\\text{NORDITA 2023-063}\end{array}$}

\begin{document}
\maketitle
\flushbottom
\setcounter{page}{2}

\section{Introduction}\label{sec:intro}

\begin{figure}[t]
	\begin{center}
		\begin{subfigure}{.48\textwidth}
			\includegraphics[width=\textwidth]{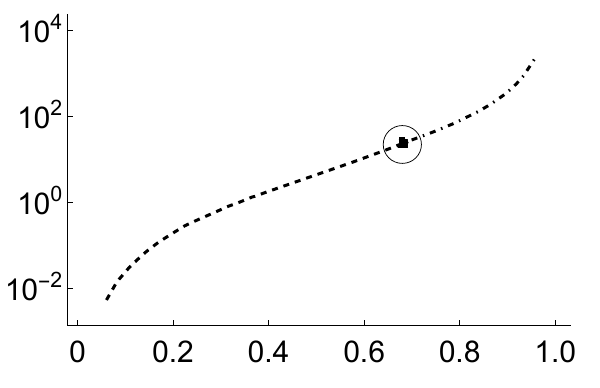} 
			\put(-175,120){\rotatebox{0}{$ T_c / \Ltwo $}}
			\put(-17,30){$b_0$}
            \put(-160,90){\footnotesize \textit{plasma phase}}
            \put(-100,50){\footnotesize \textit{gapped phase}}
		\end{subfigure}\hfill
		\begin{subfigure}{.48\textwidth}
			\includegraphics[width=\textwidth]{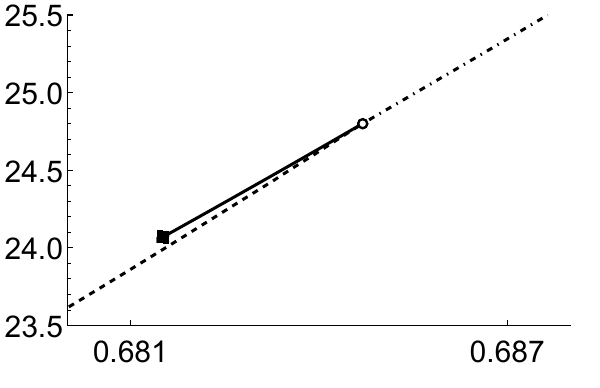} 
			\put(-175,120){\rotatebox{0}{$ T_c / \Ltwo$}}
			\put(-17,30){$b_0$}
            \put(-160,90){\footnotesize \textit{plasma phase}}
            \put(-100,50){\footnotesize \textit{gapped phase}}
		\end{subfigure}
		\caption{\small Phase diagram for the entire family of gauge theories (\textbf{left}) and zoomed in version of the disk around $b_0\approx 0.68$ (\textbf{right}) that are under study in this work. The meaning of the axes is explained in Section~\ref{sec:background}. Here, the black square stands for a critical point analogous to the one expected to be found in the QCD phase diagram, at the end of a line of first-order phase transitions. The circle, on the other hand, indicates the position of a triple point, where three phases coexist.}
		\label{fig:PhaseDiagram}
	\end{center}
\end{figure}
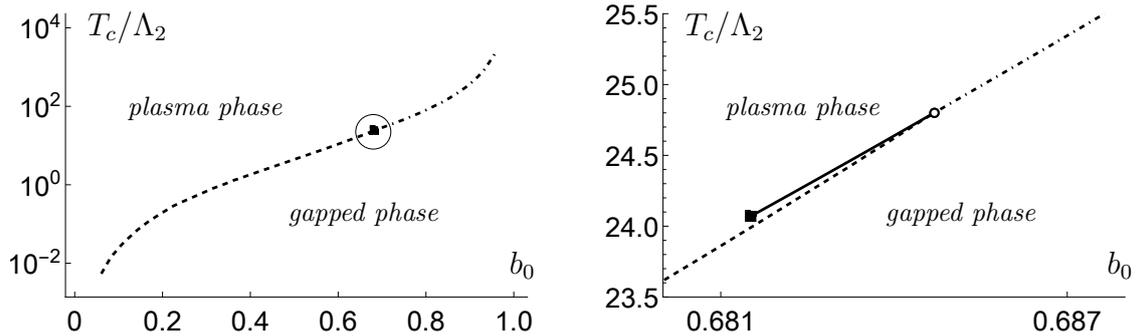

The question of whether entanglement entropy can probe different phases of a given system has a venerable history. Starting in Ref.~\cite{Klebanov:2007ws}, hope there was that the emergence of  {\emph{sharp}} phase transitions in some entanglement measures as the entangling region varies could probe the confining nature of the vacuum of a given theory. Unsettlingly, counterexamples to this statement can be found: either because the entanglement phase transition is not realized in some confining background \cite{Kol:2014nqa,Nunez:2023nnl} or because a phase transition is found in non-confining backgrounds \cite{Jokela:2020wgs,Jokela:2019tsb}. While all these works use holographic entanglement entropy (HEE) for subsystems in their analysis, other interesting takes exist in the context of (weakly coupled) algebraic quantum field theory, by means of relative entropy computations for regions with non-trivial topology \cite{Casini:2021tax} and in the context of quantum circuits \cite{Moghaddam:2023yll}. Exploration of phase transitions in entanglement measures is also entrenched in lattice simulations for gauge theories possessing confining vacua \cite{Velytsky:2009wl,Buividovich:2008kq,Nakagawa:2009jk,Nakagawa:2010kjk,Itou:2015cyu,Rabenstein:2018bri,Rindlisbacher:2022bhe,Bulgarelli:2023ofi,Jokela:2023yun,Bulgarelli:2023fgv}, an approach which chirps about the link between the sharp entanglement phase transition and the large-$N$ limit.

To observe a phase transition in some entanglement measure is a major undertaking, not least because the entanglement entropy itself is not an observable, but that one is instructed to vary the subsystem size. It is more natural to fix the sample and vary control parameters, such as the temperature of the heat bath or the magnitude of the external magnetic field. In fact, since the entanglement entropy, as defined by the von Neumann entropy of the reduced density matrix, matches onto the thermal entropy for large systems (see, {\emph{e.g.}}, \cite{Jokela:2023yun} and \cite{Ryu:2006bv} for HEE in particular) we bank on using entanglement measures to probe phases of matter also for finite and {\emph{fixed}} subsystems accompanied by the standard disclaimers on meeting requirements of the thermodynamic limit. Indeed, interesting works have investigated phases of strongly coupled systems using holographic entanglement measures \cite{Knaute:2017lll,Asadi:2022mvo,Gong:2023tbg,Zeng:2016fsb,Zhang:2016rcm} and, for example, authors of \cite{Knaute:2017lll} found a way to extract the critical exponent $\alpha$ at the critical point from a quantity inspired by the HEE. A direct computation on the lattice \cite{Jokela:2023yun} showed that approximating the entanglement entropy with the second R\'enyi entropy for 3d SU(2) indeed supplies the correct value for $\alpha$. Moreover, there are several works proposing entanglement entropy as a probe of quantum critical points in the context of applied holography for condensed matter systems (see for example Refs.~\cite{Baggioli:2020cld,Baggioli:2023ynu,Ling:2015dma,Ling:2016wyr}).

Inspired by these partial successes, we would like to explore how far can the holographic entanglement measures be stretched in the study of thermodynamic phases of strongly coupled systems. We would like to emphasize that we are not focusing on the (sharp) phase transition in the transition of the Ryu-Takayanagi (RT) surface of the HEE to another one due to altering system size but the phase transition present in the ambient field theory whose imprints we chase in HEE. To this end, we consider a specific family of gravity solutions to type IIA supergravity, called $\B_8$ class, that have duals on the field theory side. Among other interesting properties that will be reviewed later, the field theories in general possess a mass gap but are not confining, except at a specific point in the parameter space. Actually, it is in this setup that some of us showed in Ref.~\cite{Jokela:2020wgs} that, at vanishing temperature, the HEE cannot distinguish between confining and non-confining theories. 

Remarkably, in Ref.~\cite{Elander:2020rgv} is was shown that when this specific $\B_8$ class is heated up, a rich phase diagram emerges, instituting a wonderful framework for our investigations. As seen in Fig.~\ref{fig:PhaseDiagram}, the phase diagram contains three types of phase transitions, whose nature we will also review. Interestingly, the phase diagram is endowed with a critical point reminiscent to the one expected to be present in the QCD phase diagram. The goal of this paper is to extend the previous study \cite{Jokela:2020wgs} to the finite temperature case in order to understand how much information rooted in the different types of phase transitions can be extracted from entanglement considerations.

The paper is organized as follows. In Section \ref{sec:background} we review the background solutions and explain their properties; both at zero and finite temperature. In Section~\ref{sec:EE} we define several quantities, that we will use to probe the phase transitions of the background. In Section~\ref{sec:results} we analyze the lessons that we learn when we apply the general formalism to our solutions. We conclude with a discussion of the results in Section~\ref{sec:discussion} and make some general comments on the applicability of entanglement measures in the description of strongly coupled systems at large-$N$. We also comment on possible extensions of our work. Multiple appendices gather technicalities of our computations.

\section{Background solutions}\label{sec:background}

Our starting point are black brane solutions constructed in Ref.~\cite{Elander:2020rgv}, which we review in this section. These background solutions lead to a very rich phase structure. This fact constitutes a unique arena where questions regarding the capability of entanglement entropy to probe phase transitions can be addressed properly.

\subsection{Supersymmetric ground states}

Before discussing the finite temperature states of the theories, let us understand the ground states of the system at vanishing temperature. These are described holographically by the family of solutions to eleven-dimensional supergravity studied in Ref.~\cite{Faedo:2017fbv}. They constitute a one-parameter family of solutions, sourced by a stack of $N$ coincident M2-branes with an eight-dimensional transverse space from the $\B_8$ class, originally found in Refs.~\cite{Bryand:1989mv,Gibbons:1989er,Cvetic:2001ma,Cvetic:2001ye,Cvetic:2001pga}. They have Spin(7) holonomy and so they preserve $\NN = 1$ supersymmetry. There is a four-cycle whose size remains finite at the origin, while the rest of the compact part of the geometry collapses smoothly. This mechanism, as in the Klebanov--Strassler background \cite{Klebanov:2000hb} or in the Witten soliton geometry \cite{Witten:1998zw}, introduces a finite energy scale in the dual theory. 

Interestingly, $\B_8$ theories do not confine in general, {\emph{i.e.}}, the quark-antiquark potential do not show a linear growth at large separation \cite{Faedo:2017fbv}. From the gauge theory perspective, this is probably a consequence of the presence of Chern--Simons (CS) interactions in the theory. Indeed,  it is well known that, in three dimensions, gauge bosons acquire a mass in the presence of CS interactions. For this reason, color charges are screened and flux tubes between quarks can break. Geometrically this is realized by the collapse of the M-theory circle in eleven dimensions (see the argument in Ref.~\cite{Faedo:2017fbv}).
  
\begin{figure}
	\centering
	\begin{tikzpicture}[scale=3.3,very thick,decoration={markings,mark=at position .5 with {\arrow{stealth}}}]
	\node[above] at (0,0) {SYM-CSM $|$ D2};
	\node[below] at (0,-2.2) {Mass gap $|$ $\mathbb{R}^4\times {\rm S}^4$};
	\node[left,red] at (-1,-1) {OP $|$ CFT};
	\node[right] at (2,-2) {Confinement $|$ $\mathbb{R}^3\times{\rm S}^1\times {\rm S}^4$};
	\draw[postaction={decorate},ultra thick,c4] (0,0) --  (0,-2) node[left,midway]{$\mathbb{B}_8$};
	\draw[postaction={decorate},ultra thick, gray] (-1,-1) -- (-1,-2) node[left,midway]{$\mathbb{B}_8^{\textrm{\tiny OP}}$\,\,};
	\draw[postaction={decorate},ultra thick, c2] (0,0) .. controls (-.9,-.9) and (-1,-1) .. (-0.98,-2);
	\draw[postaction={decorate},ultra thick, c3] (0,0) .. controls (-.5,-.5) and (-.5,-1.5) .. (-.5,-2);
	\draw[postaction={decorate},ultra thick, c5] (0,0) .. controls (.5,-.5) and (.5,-1.5) .. (.5,-2);
	\draw[postaction={decorate},ultra thick, c6] (0,0) .. controls (.9,-.9) and (.95,-1.05) .. (1,-2);
	\draw[postaction={decorate},ultra thick, c7] (0,0) .. controls (.9,-.9) and (1.5,-1.6) .. (1.5,-2);
	\draw[postaction={decorate},ultra thick, c8] (0,0) .. controls (.95,-.95) and (1.8,-1.8) .. (1.9,-2);
	\draw[|-|] (-1,-2) -- (0,-2);
	\draw[-stealth] (0,-2) -- (2,-2);
	\node[left=5] at (-1,-2) {$b_0$};
	\node[below=5] at (-1,-2) {$0$};
	\node[below=5] at (0,-2) {$2/5$};
	\node[below=5] at (2,-2) {$1$};
	\draw[postaction={decorate},ultra thick,c1] (0,0) -- (-1,-1) node[left,midway]{$\mathbb{B}_8^\infty$\,};
	\draw[postaction={decorate},ultra thick,c9] (0,0) -- (2,-2) node[right,midway]{\, $\mathbb{B}_8^{\rm{conf}}$};
	\draw [red, ultra thick,fill=red] (-1,-1) circle [radius=0.03]; 
	\draw [black, fill=black, ultra thick] (0,0) circle [radius=0.03];
	\end{tikzpicture}
	\caption{\small Pictorial representation of the energy chart in the field theory space we study. The asymptotic UV regime is given by the 3D SYM-CSM theories. The arrows represents the renormalization group  flow from the UV to the different IR regimes. It generically drives the theory to an IR regime where a certain energy scale emerges. Only for the extreme value $b_0=1$ does the theory develop a confining behavior, as depicted on the bottom-right corner of the plot. For $b_0=0$, in contrast, the IR is governed by an Ooguri--Park conformal fixed point. The hue or the warmth of the curves will be roughly in one-to-one correspondence with the values of $b_0$ on the horizontal axis. Plot lifted from Ref.~\cite{Jokela:2020wgs}.
	}\label{fig:triangle}
\end{figure}
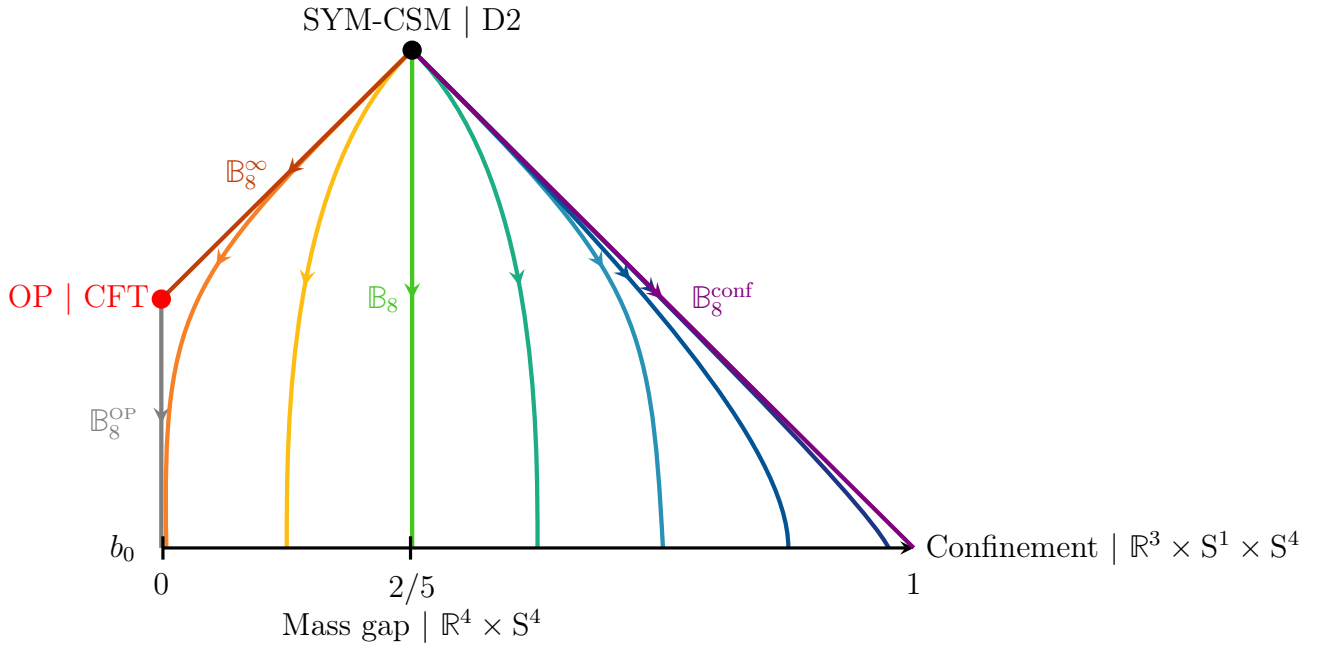

Even though the solutions we consider are regular only in eleven dimensions, their properties are better understood in terms of their reduction to type IIA supergravity. All the solutions have internal manifold of $\CP^3$. The metric ansatz in string frame is chosen so that, at the UV, the metric for a stack of D2-branes will be recovered,
\be \label{eq.metric10D}
    \dd s^2_{\text{\tiny st}} = h^{-\frac{1}{2}}\left(-\mathsf{b} \dd t^2 + \dd x_1^2+\dd x_2^2\right)+h^{1/2}\left(\frac{\dd r^2}{\mathsf{b}}+e^{2f}\dd \Omega_4^2 +e^{2g}\left[\left( E^1\right)^2 +\left(E^2\right)^2\right] \right)\,,
\ee 
with dilaton $e^\phi = h^{1/4}e^\Lambda$. In Eq.~\eqref{eq.metric10D}, $E^1$ and $E^2$ describe a two-sphere S$^2$ fibration over the four sphere S$^4$, whose volume form is $\dd \Omega_4^2$ (see Appendix~\ref{ap:solution} for more details). Moreover, we are interested in homogeneous solutions, which means that the functions $\mathsf{b}$, $h$, $f$, $g$, and $\Lambda$ will only depend on the radial coordinate $r$.\footnote{With this choice of coordinates, the boundary is approached as $r$ grows to infinity.} The supersymmetric ground states require $\mathsf{b}=1$, when Lorentz invariance is restored. However, we included the blackening factor $\mathsf{b}$ in our ansatz so that it will also describe plasma states corresponding to black brane solutions with $\mathsf{b}\neq1$, as we shall see later.

As the depiction Fig.~\ref{fig:triangle} suggests, all the backgrounds we consider share the same UV (large $r$, asymptotic) behavior. This is nothing but the one sourced by $N$ coincident D2-branes in the decoupling limit, 
\begin{equation}\label{eq:toD2intheUV}
    e^{2f} = 2 e^{2g} \sim r^2\,,\qquad e^\Phi \sim h^{\frac{1}{4}}\,, \quad h \sim N r^{-5}\,.
\end{equation}
The microscopic regime of the system is governed by a type of super Yang-Mills theory with (dimensionful) gauge coupling $\lambda = \ls^ {-1} \gs N$, with string length $\ls$, and string coupling constant $\gs$. Notice that, asymptotically, the internal manifold is described by the squashed Fubini--Study metric on $\CP^3$, and for this reason $e^{2f}/e^{2g} = 2$ (nearly K\"ahler point of $\CP^3$). On top of that, due to the different radial dependence of $f$ and $g$, this squashing changes along the flow.

Following Ref.~\cite{Loewy:2002hu}, the gauge theory dual consists of a two-site Yang-Mills quiver U($N$)$\times$U($N$) gauge group and bifundamental matter, very much like the Klebanov--Witten (KW) quiver in four dimensions \cite{Klebanov:1998hh}. Moreover, the system at hand is endowed with a non-vanishing two-form
\begin{equation}
    F_2 = Q_k J_{\rm K}\,,
\end{equation}
with $J_{\rm K}$ the K\"ahler form of $\CP^3$ and $Q_k = \ls \gs k /2$ a negative\footnote{Our system admits solutions in which $Q_k >0$ which we do not discuss.} constant with dimensions of length. Here $k$ is the  Chern--Simons level. Thus, this two-form induces CS interactions in the gauge theory dual. Additional three- and four-form fluxes manifest the presence of fractional D2-branes, which are expected to introduce a shift in the rank of one of the two gauge groups \cite{Aharony:2008gk}.

In conclusion, these gravity solutions are conjectured to describe renormalization group (RG) flows in a 
\begin{equation}
    \text{U($N$)$_k\times$U($N+M$)$_{-k}$}
\end{equation}
quiver gauge theory with CS interactions at level $k$ while preserving $\NN = 1$ supersymmetry.

As mentioned, there is a one-parameter $b_0$ family of supergravity solutions. This parameter is proportional to the asymptotic value of the Neveu--Schwarz (NS) form, and it is therefore interpreted as the difference between the microscopic Yang--Mills couplings of each of the two factors in the gauge groups,
\begin{equation}\label{eq:b0.coupligs}
    b_0 \sim \frac{1}{g_1^2} - \frac{1}{g_2^2}\,.
\end{equation}

Following the conventions in Ref.~\cite{Faedo:2017fbv}, this quantity takes values in the interval $b_0\in[0,1]$, and it allows us to represent the whole family of solutions as in Fig.~\ref{fig:triangle}. For a generic choice of the parameter $b_0\in (0,1)$, the theory does indeed develop a mass gap as we flow to the IR.\footnote{For instance, the spectrum of spin-0 and spin-2 fluctuations were computed in Ref.~\cite{Elander:2018gte}.} The two limiting values are, however, special. When $b_0 = 0$ the gap is lost and the theory flows in the IR to the Ooguri--Park (OP) CFT \cite{Ooguri:2008dk}, which is a deformation of the ABJM theory \cite{Aharony:2008ug} preserving $\NN=1$ supersymmetry. In the opposite limit, $b_0=1$, not only does the theory possess a mass gap but it also becomes confining. In fact, in this limit the CS interactions vanish and confinement is therefore expected. In contrast to the $b_0 \neq 1$ case, the way this is realized geometrically is that, for this particular case, the M-theory circle does not collapse at the IR anymore. The particular expressions for the solutions for every case can be found in Ref.~\cite{Faedo:2017fbv}.

It is useful to define two energy scales in terms of gauge theory parameters as follows
\begin{equation}\label{eq:units}
    \Lone = \frac{\lambda}{8 N}\left(\frac{|k|^ 5}{12\pi^4}\right)^{\frac{1}{3}}\,,\qquad \Ltwo = \frac{k^2 \lambda}{6\pi N}\cdot\frac{1}{(\overline M^2 + 2 |k| N)^ {\frac{1}{2}}}\,,
\end{equation}
where $\overline{M} = M-k/2$. They control the different scales at which the IR scale and/or the CS interactions become important. Following Ref.~\cite{Hashimoto:2010bq}, one should in principle be able to relate them to the two different dimensionful gauge couplings $g_1$ and $g_2$ in Eq.~\eqref{eq:b0.coupligs}, as we comment on at the end of Appendix~\ref{ap:solution}. We will use $\Lambda_1$ and $\Lambda_2$ to set the units of the different quantities that we will treat. 

Now that we summarized the main features of the zero temperature ground state, let us consider finite temperature solutions, which will be the main focus of our work.

\subsection{Low temperature states: gapped phase}

From the zero temperature solutions we have just introduced, it is straightforward to construct thermal solutions that will be the dominant ones at low temperatures and continuously connected with the supersymmetric ones discussed above. Indeed, this extension to finite temperature is obtained by going to Euclidean space and compactifying the time direction on a circle. As usual, the period $\beta$ of the Euclidean time is related to the temperature $T$ in the field theory side as 
\begin{equation}
    \beta = T^{-1} \ .
\end{equation}

These solutions exist at any temperature, since nothing is fixing the period $\beta$. The free-energy is the same as that of the ground state and independent of the temperature due large-$N$. Consequently, the entropy density is zero, as it has to be since no horizon is present in the geometry. These low temperature states will compete with a plasma phase at higher temperatures dual to black brane solutions, which we discuss next.

\subsection{High temperature states: plasma phase}\label{sec:thermo}

The high temperature phases of the system are described by black brane solutions, constructed numerically originally in Ref.~\cite{Elander:2020rgv}. For this problem the shooting method turned out to be a good procedure. With this approach, the unknown functions are solved perturbatively both about the UV region and the horizon. Next, these expansions serve as the boundary conditions used to solve the equations of motion numerically. The equations are solved starting at the two end points of the domain up to an intermediate point. In this way, the value of the parameters can be adjusted so that the functions are continuous and differentiable at the matching point.

The boundary expansions of the metric and the dilaton take the form
\begin{equation}\label{eq:UVexp}
    \begin{aligned}
        e^{f} &= \frac{|Q_k|}{u\sqrt{2}}\left(1 + \ldots + f_4 u^4 + f_5 u^ 5 + \ldots \right) \,,\qquad  e^{g} = \frac{|Q_k|}{2u}+\ldots\,,\\
        e^\Lambda &= 1+\ldots \,, \qquad \mathsf{b} = 1+ \mathsf{b}_5 u^ 5 + \ldots \,, \qquad h = \frac{4q_c^2 + 3Q_c |Q_k|}{|Q_k|^6}  \frac{16}{15}(1- b_0^2) u^ 5+\ldots\,    
    \end{aligned}
\end{equation}
where we changed to the radial coordinate $u = |Q_k|/r$. Here $Q_c$ and $q_c$ are dimensionful quantities related to gauge theory quantities in the way explained in Appendix~\ref{ap:solution}. For our purposes it is enough to display only those coefficients in Eq.~\eqref{eq:UVexp} that enter in the expressions for the thermodynamic quantities appearing later (see Ref.~\cite{Elander:2020rgv} for further details).

The existence of a horizon is encoded in a simple zero of the blackening factor $\mathsf{b}$ at some value of the radial coordinate $u=u_h$. The leading terms in the expansion of the metric and the dilaton about the black brane horizon, which contain the undetermined parameters entering in the thermodynamic expressions, are
\begin{equation}
    \begin{aligned}
        e^f &= |Q_k| f_h + \ldots\ , \qquad e^g = |Q_k| g_h + \ldots\ , \quad e^\Lambda = \lambda_h + \ldots\ \\
    h &= \frac{4q_c^2 + 3Q_c |Q_k|}{|Q_k|^6} h_h + \ldots\ , \qquad \mathsf{b} = \mathsf{b}_h (u-u_h) + \ldots \ .
    \end{aligned}
\end{equation}
The period of the Euclidean time (after imposing regularity) and the area density of the black brane horizon lead, respectively, to expressions for the temperature $T$ and entropy density $S$ of the field theory, namely \begin{equation}\label{eq:entropy_temperature}
    T = \frac{\Ltwo}{4\pi}\cdot\frac{(-\mathsf{b}_h) u_h^2}{\sqrt{h_h}}\,,\qquad
    S = \frac{\Lone^ 3}{\Ltwo}\cdot \frac{64\pi f_h^4 g_h^2 \sqrt{h_h}}{\lambda_h^2}\,.
\end{equation}
The free energy density is obtained from the on-shell four-dimensional bulk action, given in terms of UV data as
\begin{equation}\label{eq:free_energy}
    F = \Lone^ 3\left(-\frac{411}{2}-6f_4 - 2f_5 + \frac{3}{2}\mathrm{b}_5\right)\,.
\end{equation}
Note that expressions in Eqs.~\eqref{eq:entropy_temperature} only depend on horizon data. In particular, the relation $S=-\dd F/\dd T$ can be used as a crosscheck for the numerics.

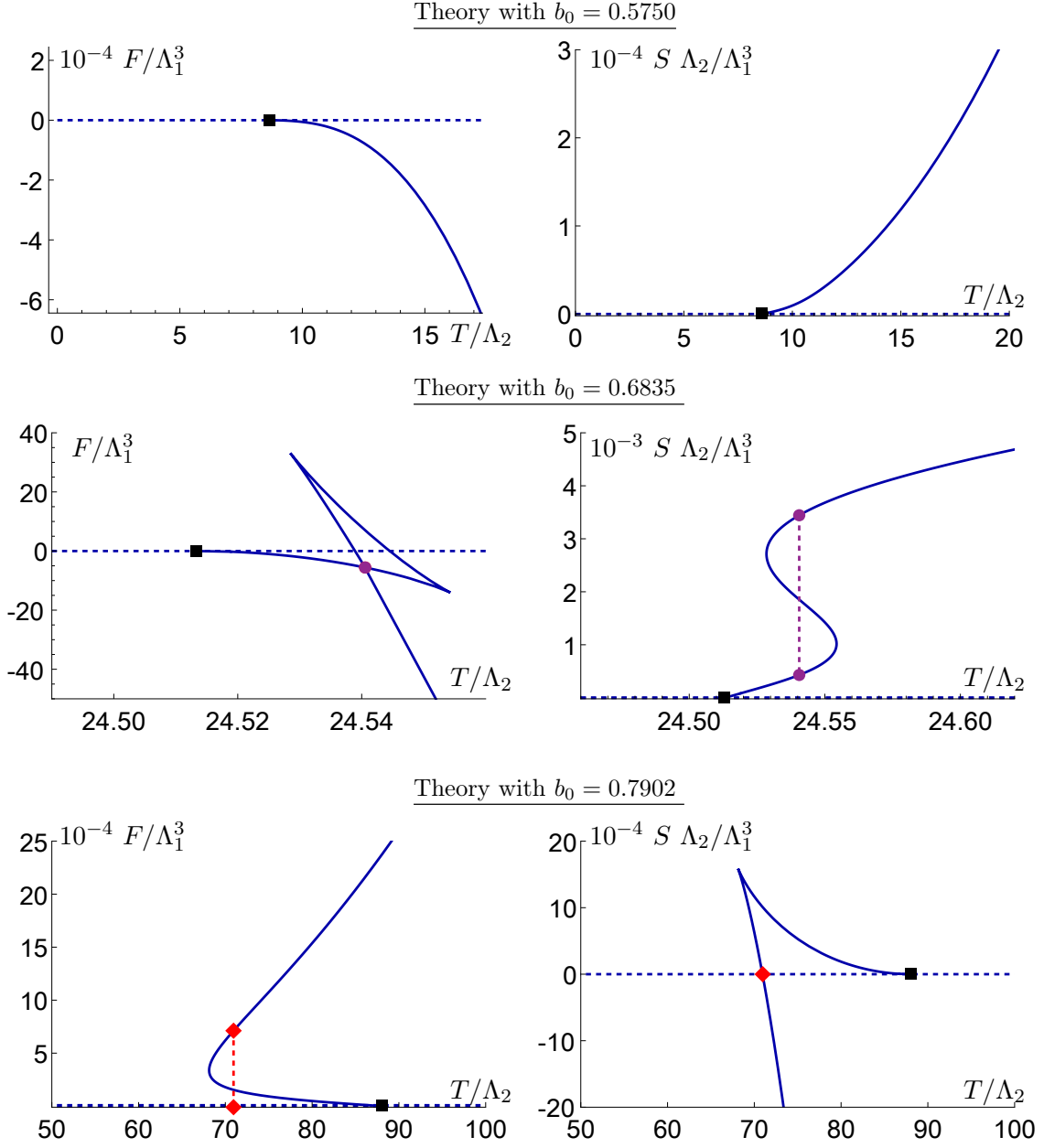
\begin{figure}[htb]
	\begin{center}
    \includegraphics[width=\textwidth]{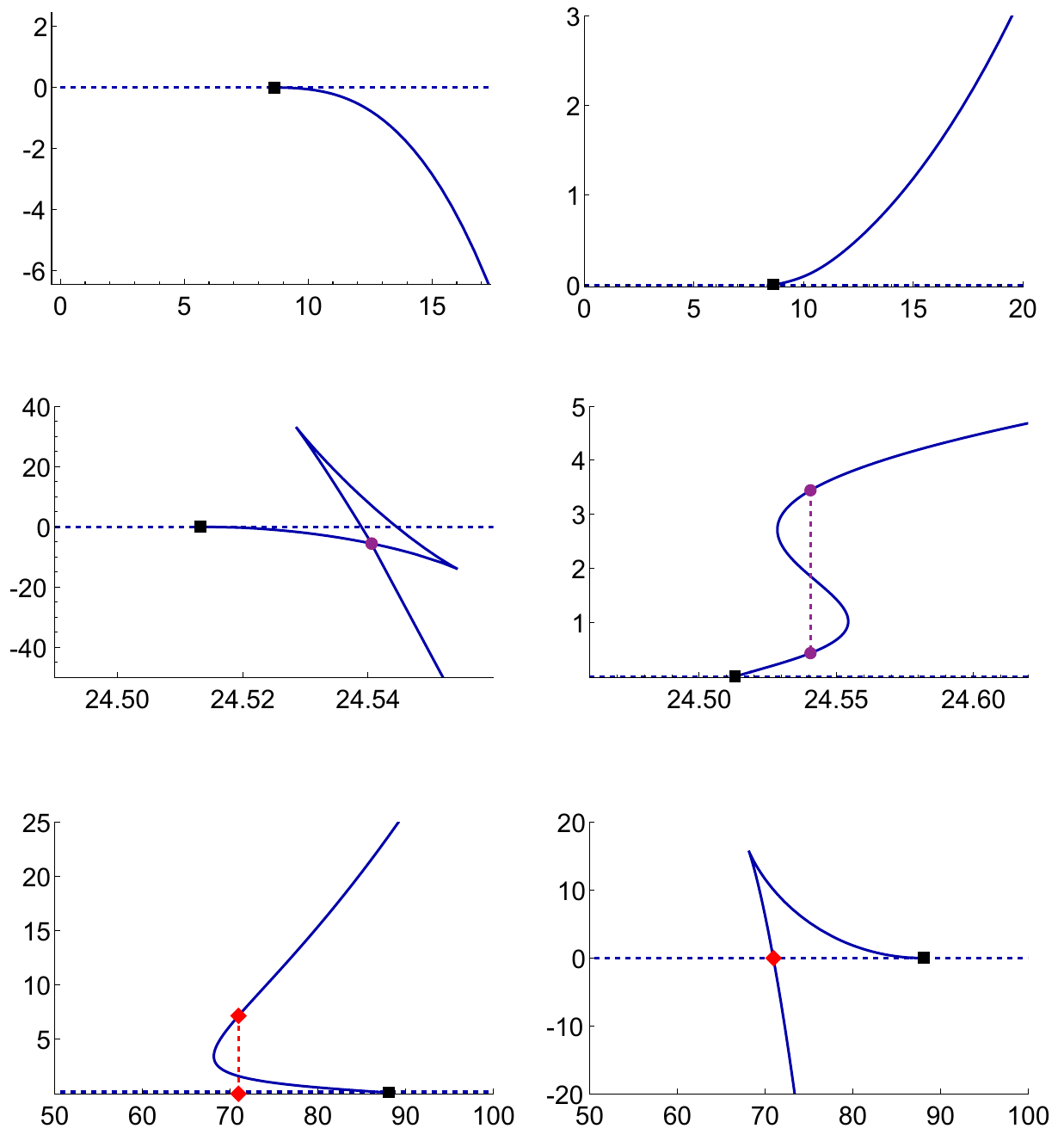} 
   \put(-260,470){\underline{\footnotesize Theory with $b_0 = 0.5750$}}
    \put(-260,315){\underline{\footnotesize Theory with $b_0 = 0.6835$ }}
    \put(-260,150){\underline{\footnotesize  Theory with $b_0 = 0.7902$ }}
    \put(-188,450){\small$ 10^{-4}\  S \ \Ltwo/\Lone^3$}
    \put(-188,290){\small$ 10^{-3}\  S \ \Ltwo/\Lone^3$}
    \put(-188,130){\small$ 10^{-4}\  S \ \Ltwo/\Lone^3$}
    \put(-405,450){\small $10^{-4}\ \small F / \Lone^3$}
    \put(-400,290){\small $\small F / \Lone^3$}
    \put(-405,130){\small $10^{-4}\ \small F / \Lone^3$}
    \put(-245,337){\small$ T /\Ltwo$}
    \put(-245,195){\small$ T/\Ltwo$}
    \put(-245,25){\small$ T/\Ltwo$}
    \put(-35,353){\small $ T/\Ltwo$}
    \put(-35,195){\small $ T/\Ltwo$}  
    \put(-35,25){\small $ T/\Ltwo$}
	\caption{\small Free energy (\textbf{left}) and entropy (\textbf{right}) densities as a function of the temperature for states in the theories with three different values of $b_0$ as representatives for the three cases discussed in the main text. Blue solid curves (dashed lines) correspond to plasma (gapped) phases. Second-order PTs between them at zero entropy but finite temperature (\textit{type~1}) are depicted using black squares. A first-order PT between plasma phases (\textit{type~2}) is represented by purple dots and a dashed purple line. Finally, the red diamonds and the dashed red line stand for a first-order PT between black brane solutions and the ground state, with a jump in the entropy (\textit{type~3}).}\label{fig.DifferentCasesThermalPT}
	\end{center}
\end{figure}

Now that we have the expressions for the thermodynamic quantities, the next step is to understand, for a each value of $b_0$, which solution is thermodynamically preferred at any given temperature. Remarkably, a rich phase structure emerges, which allows us to distinguish three different cases, depending on the choice of $b_0$. The two particular values for which the qualitative behavior changes are 
\begin{equation}
    \bcritical \approx 0.6815 \quad \text{and} \quad \btriple \approx 0.6847\,.
\end{equation}
The main features of each case can be contemplated in Fig.~\ref{fig.DifferentCasesThermalPT} and are pronounced next. We are specifically interested in the type of phase transition (PT) between the different phases.

\begin{description}
\item[$\blacksquare$] For small values of $b_0$, in the range $0 < b_0 < \bcritical$, both the free energy and the entropy densities touch zero for some finite value of the temperature. We show a representative of this case in Fig.~\ref{fig.DifferentCasesThermalPT}~(\textbf{top}). This is a peculiar kind of phase transition which could be very well claimed to be a second-order phase transition since these thermodynamic quantities are continuous but the derivative of the entropy is not.\footnote{Note, however, that in Ref.~\cite{Elander:2020rgv} the fact that the geometry jumps discontinuously at that point made the authors claim that it is first order, since one would expect that some $n$-point function will be discontinuous.} Such sort of phase transition has been found on other systems, see for instance Ref.~\cite{Bena:2018vtu}. We will refer to this kind of phase transition as \textit{type 1}, characterized by the fact that the entropy raises smoothly from zero when the critical temperature is reached.
\item[$\blacksquare$] For theories in the range $ \bcritical < b_0 < \btriple$ something interesting happens: new branches of black brane solutions appear when the entropy is small enough. The plot of the free energy as a function of temperature develops a swallow-tail shape, characteristic of first-order phase transitions (see Fig.~\ref{fig.DifferentCasesThermalPT} (\textbf{middle}, \textbf{left})). Indeed, there are two locally stable branches of black hole solutions that compete and their crossing signals a first-order phase transition between two plasma phases. Note the discontinuous jump in the entropy density at this particular temperature. We will refer to this kind of phase transition as \textit{type~2}.
Note that, for the particular choice of $b_0$ shown in the Fig.~\ref{fig.DifferentCasesThermalPT} (\textbf{middle}), at a lower temperature we still find a \textit{type 1} phase transition where the entropy vanishes at finite temperature.
\item[$\blacksquare$] When $\btriple < b_0 < 1$, the \textit{type 1} phase transition is hidden below a branch of stable black brane solutions, as seen in Fig.~\ref{fig.DifferentCasesThermalPT} (\textbf{bottom}). When this happens, the \textit{type~2} phase transition is also lost. The situation is reminiscent of the Hawking--Page phase transition \cite{Hawking:1982dh}: the free energy of the branch of black brane solutions crosses the horizontal line at some critical temperature $T_c$, where a first-order PT to the ground state occurs. The fact that the entropy density jumps discontinuously at $T_c$ is again a manifestation of the transition being first order. Note that this is a ``degapping'' phase transition rather than a deconfinement phase transition, for the low temperature ground state is not confining. We refer to this kind of phase transition as \textit{type 3}.
\item[$\blacksquare$] Finally, let us comment on the limiting values of $b_0$. For $b_0 = 0$ no phase transition is present. The reason is that the theory flows to a CFT at low temperatures, where all the quantities develop conformal behavior ($S\propto T^2$, for example). Put differently, as $b_0\to0$ the critical temperature corresponding to \textit{type 1} phase transition approaches zero. For $b_0=1$ the ground state corresponds to a truly confining theory, and the \textit{type~2} phase transition becomes a genuine deconfinement phase transition in this case. Additionally, as $b_0\to1$, the temperature at which entropy vanishes goes to infinity. These distinct limiting behaviors are manifest in Fig.~\ref{fig:PhaseDiagram}.
\end{description}

In conclusion, there are three distinct types of phase transitions that we encounter in these theories. Let us examine how, and how well, the measures of entanglement entropy probe them.

\newpage

\pagebreak

\section{Entanglement thermodynamics}\label{sec:EE}

In this section, we will define several quantities related to entanglement entropy that will provide some information about the thermal phase diagram of the system. Then, these quantities will be applied to our case in the subsequent Section~\ref{sec:results}.
 
\subsection{Entanglement entropy}

Following Ref.~\cite{Ryu:2006bv}, the entanglement entropy of a QFT region $A$ bounded by $\partial A$ is given holographically by the area of a minimal surface $\Sigma_A$ anchored on $\partial A$ at the boundary of spacetime. The minimal surface $\Sigma_A$ is also homologous to $A$. This minimal surface $\Sigma_A$ has co-dimension two, {\emph{i.e.}}, it is an eight-dimensional submanifold embedded in the ten-dimensional background geometry \eqref{eq.metric10D}. This surface wraps completely the compact part of the geometry and has a particular profile along the radial direction. We will refer to it as the RT surface associated to $A$.  

Thus, the holographic entanglement entropy in string frame reads
\be \label{eq:actionSigma}
    \SEE_A = \frac{1}{4G_{10}}\int_{\Sigma_A} \dd^8\sigma e^{-2\Phi}\sqrt{\det \mathsf{g}}\ ,
\ee 
where the $\sigma$’s are coordinates on $\Sigma_A$, the ten-dimensional Newton's constant reads $G_{10} = (16\pi)^{-1}
(2\pi)^7g_s^2\ell ^8_s$, the function $\Phi$ is the dilaton, and $\mathsf{g}$ is the induced metric on the surface in string frame,
\begin{equation}\label{eq:induced-metric}
    \mathsf{g}_{\alpha\beta} = \frac{\partial x^\mu}{\partial \sigma^\alpha}\frac{\partial x^\nu}{\partial \sigma^\beta}g_{\mu\nu}\,,
\end{equation}
with $g_{\mu\nu}$ given by Eq.~\eqref{eq.metric10D}.
The surface $\Sigma_A$ is wrapping the whole compact internal manifold, so integration over the six corresponding coordinates gives a factor $V_6 = 32\pi^3/3$, which is nothing but the volume of $\CP^3$. The RT surface is prescribed to be static, thus the embedding is determined by
\begin{equation}
    t = \text{constant\ ,}\qquad 
    x_1 = x_1(\sigma_1,\sigma_2)\ ,\qquad 
    x_2 = x_2(\sigma_1,\sigma_2)\ ,\qquad 
    r = r(\sigma_1,\sigma_2)\ .
\end{equation}

Varying the action in Eq.~\eqref{eq:actionSigma} with respect to these fields we obtain the Euler--Lagrange equations
\be 
    \frac{\pat \mathcal{L}}{\pat \phi^i} - \pat_\mu\left ( \frac{\pat \mathcal{L}}{\pat(\pat_\mu \phi^i)}\right) = 0\ , \qquad \text{for }\phi^i \in \{x_1,x_2,r\} \text{ and } \mu = \sigma_1,\sigma_2\ ,\label{eq:Euler-Lagrange-general}
\ee 
where $\mathcal{L}$ is just the integrand in Eq.~\eqref{eq:actionSigma}. In the general case, Eq.~\eqref{eq:Euler-Lagrange-general} leads to a system of three second-order partial differential equations, which are the equations that the embedding has to fulfill so that it is extremal. In this paper, we will only analyze entanglement entropy of (infinite) strips. We believe that this analysis will capture the main features when in comes to understanding how entanglement entropy probes finite temperature states of strongly-coupled quantum field theories.

Let us then take the region $A$ to be a strip of width $l$. In contrast to the zero temperature case (see Fig.~3 in Ref.~\cite{Jokela:2020wgs}), whenever a deconfined phase (dual to a black brane solution) is considered, the only relevant configurations are the ``connected'' ones, which we will denote by~$\cup$. More precisely, in gapped theories at zero temperature there is another extremal configuration that competes with the ``connected'' one and eventually becomes the dominant one. This consists of two pieces extending vertically from the UV down to the IR plus a piece that lies at the bottom of the geometry. Crucially, the second piece has zero area and does not contribute to the entanglement entropy. In contrast, in the presence of a horizon, a surface that lies on the horizon would actually contribute, with the ultimate consequence that the ``disconnected'' configuration is never realized.

As a consequence, we only consider ``connected'' configurations, denoted by $\cup$ and specified by the choice
\be 
    t = \text{constant}, \quad  x^1 = \sigma^1 \in [-l/2,l/2], \quad  x^2 = \sigma^2 \in \R, \quad  r = r(\sigma^1) \in [r_*,\infty) \ ,
\ee 
where $r_* (> r_H)$ is the value of the radial coordinate at which the RT surface has a turning point. This choice reduces the system of Eqs.~\eqref{eq:Euler-Lagrange-general} to a single second-order ordinary differential equation. Furthermore, the expression for the entanglement entropy in this case reads
\begin{equation}\label{eq:EEconnectedConf}
    \SEE_\cup(b_0,T,l) = \frac{V_6 L_y}{4 G_{10}}\int_{-\frac{l}{2}}^ {\frac{l}{2}}\dd \sigma^1 \,   \Xi^{\frac{1}{2}} \left[1+\frac{h}{\mathsf{b}}\dot{r}^2\right]^{\frac{1}{2}}\,,
\end{equation}
where we have decided to maintain the dependence on the particular theory and the temperature explicit. In Eq.~\eqref{eq:EEconnectedConf}, the dot stands for differentiation with respect to $\sigma^1$, $L_y = \int_\R\dd \sigma^2$ is the full (infinite) integration over $\sigma^2$ and
\begin{equation}\label{eq:Xi}
    \Xi = h^2 e^{8f+4g-4\Phi}
\end{equation}
is a combination of metric functions and the dilaton that we define for convenience. 

Note that the integrand in Eq.~\eqref{eq:EEconnectedConf} does not depend explicitly on $r$. As a consequence, there is a conserved quantity that simplifies the resolution of the remaining second-order differential equation. Moreover, the result from Eq.~\eqref{eq:EEconnectedConf} is UV divergent and, if we want to compare this quantity in different thermal states, we need to regularize it not only in a width- but also temperature-independent way. This is done by introducing appropriate counterterms
\begin{equation}
    \label{eq:regularised_EE}
    \SEE_\cup^{\text{\tiny reg}} (b_0,T,l) = \SEE_\cup(b_0,T,l) - \SEE_{\text{\tiny ct}}(b_0)\,.
\end{equation}
All the details regarding the computation of $\SEE_\cup$ and its regularization are discussed in Appendix~\ref{ap:strip}. Most of the quantities that we consider are renormalization-scheme independent, as we will stress in the relevant cases.

The first observation we would like to make is that the behavior of the entanglement entropy as a function of the width of the strip looks qualitatively the same, quite generally, whenever there is a horizon present in the geometry. This is a well-known fact: the minimal surface finds it advantageous to place most of its volume at the IR, lying very near the horizon. Consequently, for large strip widths, the entanglement entropy ends up growing linearly as
\begin{equation}
    \label{eq:EE_linear_growth}
    \SEE_\cup^{\text{\tiny reg}} \propto \,  \Xi_*^{\frac{1}{2}} \, L_y\cdot l \propto S\,   L_y \cdot l\,,
\end{equation}
where $S$ corresponds to the thermal entropy density, given by Eq.~\eqref{eq:entropy_temperature} in the present case. This is the expected volume law for entanglement entropy at finite temperature.\footnote{Recall we are in $2+1$ dimensions.} Indeed, we see that for wide strips the entanglement entropy Eq.~\eqref{eq:EE_linear_growth} becomes the area of the strip times the entropy density of the corresponding black brane solution.
Note that the square root of (\ref{eq:Xi}) matches onto the thermal entropy of the field theory in the high temperature phase when evaluated at the tip of the RT surface \cite{Jokela:2020wgs}. This is by construction: the RT surface sweeps the horizon with only negligible contributions from the straight pieces stretching between the boundary and the horizon in this limit. This fact is most easily recovered by the utility of the chain rule  which equates $d\SEE/dl\propto \Xi_*^{1/2}$ \cite{Bilson:2010ff,Jokela:2020auu}.

\begin{figure}[t]
	\begin{center}
		\begin{subfigure}{0.47\textwidth}
			\includegraphics[width=\textwidth]{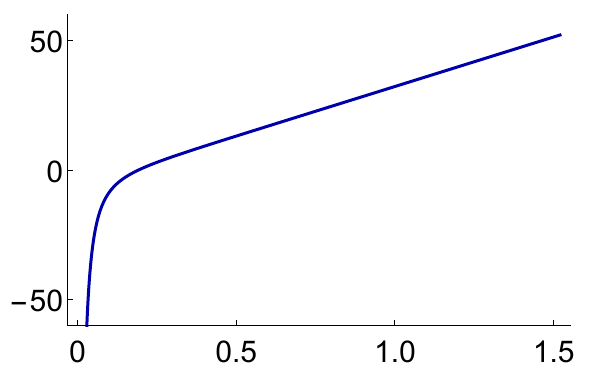} 
			\put(-170,115){\small $\SEE_\cup^{\text{\tiny reg}} \cdot \Ltwo^2 / (64\pi L_y\Lone^3)$}
			\put(-30,30){$l/l_c$}
		\end{subfigure}\hfill
		\begin{subfigure}{.47\textwidth}
			\includegraphics[width=\textwidth]{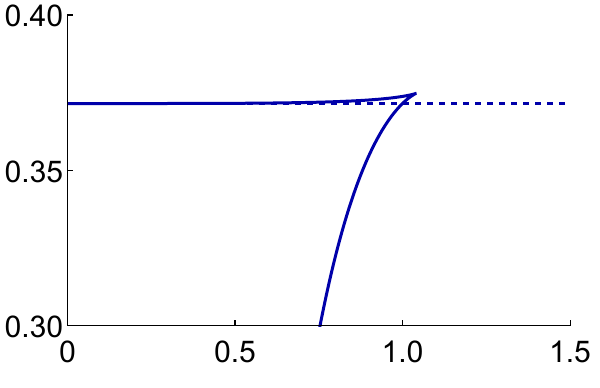} 
			\put(-175,115){ \small $\SEE_\cup^{\text{\tiny reg}} \cdot \Ltwo^2 / (64\pi L_y\Lone^3)$}
			\put(-30,30){$l/l_c$}
		\end{subfigure}
		\caption{\small We depict the holographic entanglement entropy of a single strip as a function of its width for a theory with $b_0 = 2/5$; (\textbf{left}) in a plasma phase at a temperature \mbox{$T\simeq22.78\Ltwo>T_c$} and (\textbf{right}) at in the low temperature gapped phases, for any $T<T_c$. While at finite temperature the entropy grows monotonically as a function of the width, at zero temperature a phase transition to a ``disconnected'' configuration (dashed line) is found. Here $l_c$ stands for the value at which this phase transition occurs and the widths in both plots are normalized to it.
		}\label{fig.EEplotsStrip}
	\end{center}
\end{figure}

On the other hand, the scaling for small widths is dictated by D2-brane asymptotics 
\begin{equation}
    \label{eq:EE_D2}
    \SEE_\cup^{\text{\tiny reg}} \propto - l^{-\frac{4}{3}}\,.
\end{equation}
These two facts generically lead to a smooth monotonically increasing function,\footnote{We found some richer cases where several embeddings exist for a given strip width, which would lead to the appearance of a swallow-tale shape in the $\SEE_\cup^{\text{\tiny reg}}(l)$ curve and a cusp for the corresponding preferred configuration. These, however, appear only for some black brane solutions lying on a thermodynamically unstable branch and for that reason we will not discuss them any longer here.} as shown in Fig.~\ref{fig.EEplotsStrip} (\textbf{left}).

Finally, note that the monotonous growth contrasts to what is found for entanglement entropies of strips in the low energy ground state, shown in Fig.~\ref{fig.EEplotsStrip} (\textbf{right}). There, the small width behavior of Eq.~\eqref{eq:EE_D2} remains the same, as it is governed by the UV of the theory. However, above some critical width of the strip, the preferred embedding is that of a ``disconnected'' configuration, signaling the presence of an emergent IR scale (see Ref.~\cite{Jokela:2020wgs} for details).

The fact that the behavior of entanglement entropy is always similar to the one found in Fig.~\ref{fig.EEplotsStrip} (\textbf{left}) could lead to the expectation that little information can be extracted from entanglement, when it comes to pinpointing the thermal phase structure of a theory. This is precisely the problem we address. To tackle it, we shall examine how entanglement measures vary for different theories (\textit{i.e.}, different choices of $b_0$) as we vary the temperature $T$ and the strip width $l$. We are then facing a three-dimensional parameter space. For clarity, we find it convenient to first fix the value of $b_0$, and then show how entanglement entropy varies as a function of the temperature for different choices of the width.

\subsection{Mutual information}

Note that entanglement entropy is not unambiguously defined. In particular, Eq.~\eqref{eq:regularised_EE} is scheme-dependent as one can also modify the chosen counterterms by finite quantities. In particular, it is not an observable.

For this reason, we will in some cases analyze another interesting quantity, called mutual information. The mutual information between two entangling strips $A$ and $B$, given by
\begin{equation}\label{eq:mutualinfo}
    I(A,B) = \SEE_A + \SEE_B - \SEE_{A\cup B}
\end{equation}
characterizes the amount of information shared by the two domains \cite{Headrick:2010zt}. We will study mutual information between two strips of the same width $l$ that are separated by a distance $s$. In this case Eq.~\eqref{eq:mutualinfo} can in our holographic setting be rewritten using Eq.~\eqref{eq:regularised_EE} as
\begin{eqnarray}\label{eq:mutualinfostrips}
    I(b_0,T,l,s) = 2 \SEE_\cup^{\text{\tiny reg}} (b_0,T,l) -\SEE_\cup^{\text{\tiny reg}} (b_0,T,2l+s) - \SEE_\cup^{\text{\tiny reg}} (b_0,T,s) \,.
\end{eqnarray}
Note that the counterterms drop out from this last expression when homogeneity and isotropy is assumed, manifesting that mutual information is scheme-independent.

\subsection{Entanglement pressure}\label{sec:pressure}

Now that we know how to compute entanglement entropies of strips, we shall investigate what properties of the corresponding thermodynamic phase diagram can be unveiled from entanglement considerations. The experiment we have in mind is that of placing a strip of a given width $l$ in our system, and measure how entanglement properties vary with the temperature $T$. Because we are keeping the theory labeled by $b_0$ fixed, and also the width of the strip $l$, it is useful to make this explicit by denoting the entanglement entropy in this context as $\SEE_{b_0,l}(T) = \SEE_\cup^{\text{\tiny reg}} (b_0,T,l)$.

Knowing how the strip entanglement entropy varies as the temperature changes at fixed width, we can define the corresponding \textit{entanglement pressure}, inspired by Ref.~\cite{Knaute:2017lll}, as
\begin{equation}\label{eq:def_pressure}
    P_{b_0,l}(T) - P_{b_0,l}(T_{0}) = \int_{T_{0}}^T \, \SEE_{b_0,l}(T) \, \dd T\,.
\end{equation}
Again, the subscript denotes that we are performing this integral at fixed values of $b_0$ and $l$. With this definition, we have
\begin{equation}
    \SEE_{b_0,l} = \frac{\dd P_{b_0,l}}{\dd T}\,,
\end{equation}
in analogy to the thermal case. Note that in Eq.~\eqref{eq:def_pressure} we have freedom to choose the value of the entanglement pressure at the reference temperature, $P_{b_0,l}(T_{0})$. This arbitrariness cancels out when two pressures are compared, as long as they come from integrating the entropy along the same branch of solutions.

The definition of entanglement pressure in Eq.~\eqref{eq:def_pressure} may seem physically unmotivated. In particular, it is somehow discouraging that it depends on the width of the strip, and actually this will have consequences when trying to probe some of the phase transitions presented in the Section~\ref{sec:background}. We proceed by viewing it as a way to implement Maxwell construction in the current scenario. As unsatisfactory as this may seem, we will still observe that in certain cases the answer we get (\textit{i.e.}, the value of the critical temperature) coincides with the value obtained from thermodynamics. We will elaborate on this later when we discuss each case. Let us just anticipate that with this definition it is possible to locate the critical point present in Fig.~\ref{fig:PhaseDiagram}.

\subsection{Critical exponents}

We have just mentioned that the position of the critical point can be pinpointed using the definitions that we have given. We will be interested in understanding if we can extract further properties of the critical point.

In particular, we address the question if the critical exponents can be extracted from thermodynamic quantities. For that we rely on the analogous \textit{entanglement specific heat}, defined through the logarithmic derivative of entanglement entropy with respect to temperature in the vicinity of the critical point at fixed strip width~$l$: 
\begin{equation}\label{eq:critical_exponentEE}
    C_{\bcritical,l}(T) \equiv T \frac{\dd \SEE_{\bcritical,l}}{\dd T} \sim |T-\TCP|^{-\alpha_{\text{\scalebox{.7}{EE}}}} \ .
\end{equation}

Considering the theory for which the critical phenomena is realized ({\emph{i.e.}}, $b_0 = \bcritical$), if we place ourselves near the critical point and entanglement is indeed sensitive to the critical phenomena, we expect that $C_{\bcritical,l}(T)$ develops a power law behavior. In Eq.~\eqref{eq:critical_exponentEE}, the parameter $\alphaEE$ is the corresponding critical exponent and $\TCP$ is the temperature at the critical point. Notice that in general there would be a different exponent when approaching the critical point from either low temperature ($\alphaEE$) or from high temperature ($\alphaEE'$) side. Here, we are considering field theories which are isotropic and expect to find the same exponent from both sides ($\alphaEE'\approx \alphaEE$).

In our analysis, after locating the critical point by means of entanglement quantities, we will come back to the computation of the critical exponent $\alphaEE$ and compare the result with the critical exponent obtained from thermodynamics.

\section{Probing phase transitions through entanglement}\label{sec:results}

Now we have all the ingredients to discuss how the different PTs that we encountered in Section~\ref{sec:thermo} are probed by entanglement measures. As stated in the previous section, we will discuss entanglement entropies for strips, believing that a shape dependence of the entangling region would not affect our main conclusions.

\subsection{\textit{Type 1} phase transitions}\label{sec:type1}

\begin{figure}[t]
	\begin{center}
		\begin{subfigure}{.49\textwidth}
			\includegraphics[width=\textwidth]{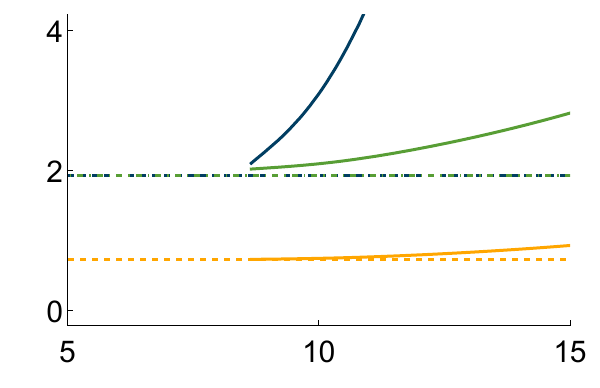} 
			\put(-195,140){\rotatebox{0}{\small $\SEE_{b_0,l}\cdot \Ltwo^2 / (64\pi L_y\Lone^3)$}}
			\put(-30,27){\small $T /\Ltwo$}
		\end{subfigure}\hfill
		\begin{subfigure}{.49\textwidth}
			\includegraphics[width=\textwidth]{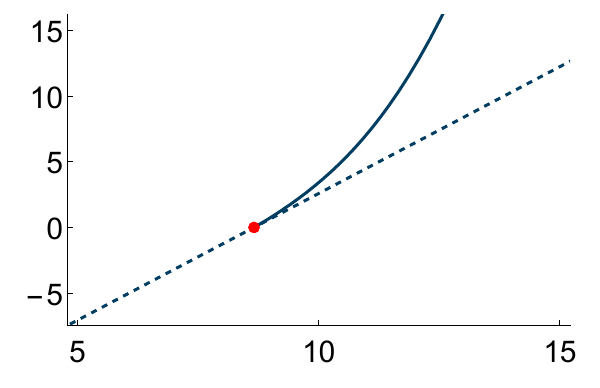}
			\put(-195,140){\rotatebox{0}{\small $P_{b_0,l}\cdot \Ltwo^3 / (64\pi L_y\Lone^3)$}}
			\put(-30,27){\small $T /\Ltwo$}
		\end{subfigure}
		\caption{\small (\textbf{Left}) Entanglement entropy as a function of the temperature for the theory with $b_0 = 0.5750$ for different fixed values of the strip width; $l = 0.01\Ltwo^{-1}$ (yellow), $0.02\Ltwo^{-1}$  (green), and $0.2\Ltwo^{-1}$ (dark blue). The dashed line stands for the  entanglement entropy for the gapped phase at the corresponding width. Note that it is the same for $l>l_c\simeq 0.184\Ltwo^{-1}$.  (\textbf{Right}) Entanglement pressure for $l = 0.2\Ltwo^{-1}$. The fact that the pressures coincide at the critical temperature is a consequence of our prescription}
		\label{fig:FixedWidth_y010}
	\end{center}
\end{figure}

Let us start by examining the \textit{type 1} phase transitions using the prescription just defined. For that, we wish to choose some particular value for the strip width and see how entanglement entropy and entanglement pressure vary as a function of the temperature. Importantly, from the point of view of the entanglement entropy, the gapped and the plasma phases are not continuously connected. Across \textit{type~1} phase transitions, the geometry changes (see Ref.~\cite{Elander:2020rgv}). This introduces a jump in the entanglement entropy of the strip, as a function of temperature, as evinced in Fig.~\ref{fig:FixedWidth_y010} (\textbf{\textbf{left}}). Still, we can attempt to compute entanglement pressure following Eq.~\eqref{eq:def_pressure}.

For the gapped phase Eq.~\eqref{eq:def_pressure} simply leads to the integration of a constant over a certain range of temperatures and thus the pressure becomes a straight line. However, we realize that the discontinuity in the entanglement entropy leads to some confusion concerning the way in which the pressure $P_{b_0,l}(T)$ of the plasma phase should be obtained. A natural choice is to prescribe that entanglement pressure must be continuous when the phase transition takes place at $T_c$. In particular, to compute the entanglement entropy for the plasma phase, we will first reach $T_c$ from some reference temperature $T_0$ after integrating the constant curve corresponding to the gapped phase. Then, at $T_c$, the integration continues along the curve obtained from the plasma phase.

In this way, the branch coming from the plasma phase starts touching the gapped phase at $T_c$, and grows afterwards above the corresponding line for the gapped phase. This is how the phase transition is revealed in this case.

Note that in our analysis the transition appears to be of first order. Indeed, entanglement entropy is discontinuous, rendering entanglement pressure continuous but non-differentiable at $T_c$. Yet, entanglement entropy is not related to any of the thermodynamic potentials, so we refrain from determining the order of the phase transition just from entanglement considerations.

In conclusion, for this type of phase transition entanglement entropy is able to locate the phase transition. However, from foresight we had already incorporated in our prescription to connect both branches of solutions (corresponding to the gapped phase and the plasma phases). The judicious choice led to the correct result. 

\subsection{\textit{Type 2} phase transitions and location of the critical point}\label{sec:type2}

\begin{figure}[t]
	\begin{center}
		\noindent \begin{subfigure}{\textwidth}
			\includegraphics[width=\textwidth]{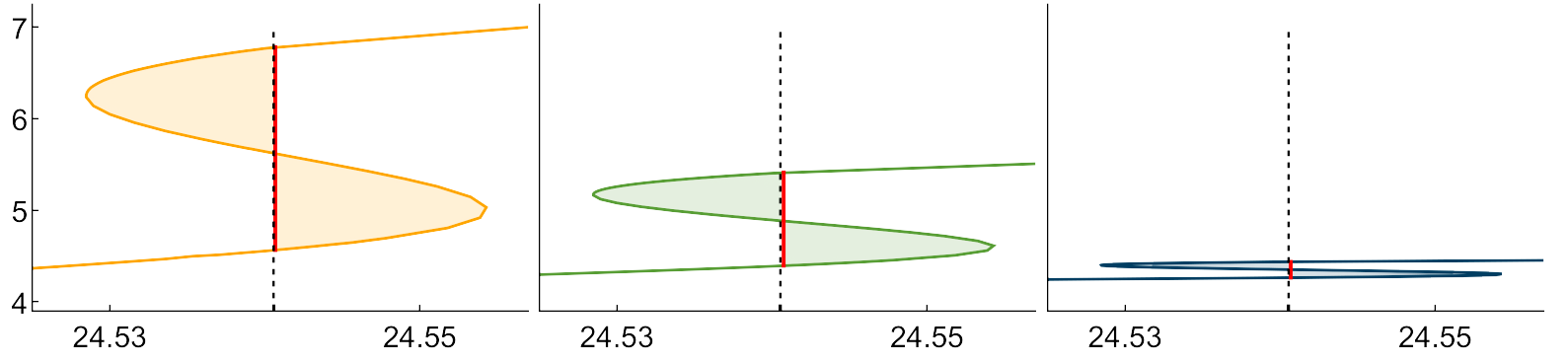}  
			\put(-430,105){\rotatebox{0}{\small $\SEE_{b_0,l} \cdot \Ltwo^2 / (64\pi L_y\Lone^3)$}}
			\put(-30,-10){\small $ T/\Ltwo$}
			\put(-310,-10){\small $T/\Ltwo$}
			\put(-170,-10){\small $T/\Ltwo$}
			\put(-65,70){\footnotesize$ l = 0.01\Ltwo^{-1}$}
			\put(-205,70){\footnotesize $ l = 0.06 \Ltwo^{-1}$}
			\put(-340,70){\footnotesize $ l = 0.14 \Ltwo^{-1}$}
		\end{subfigure}
		
		\vspace{.7cm}
		
		\noindent \begin{subfigure}{.5\textwidth}
			\includegraphics[width=\textwidth]{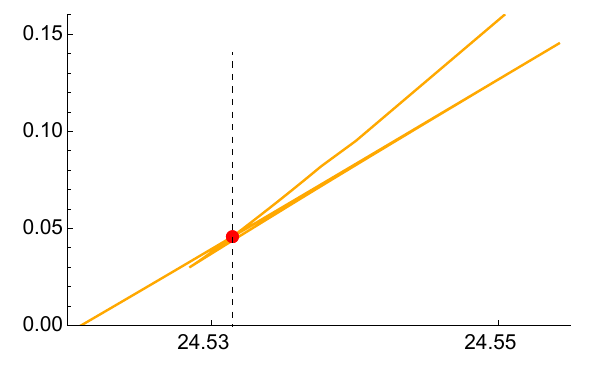} 
			\put(-205,145){\rotatebox{0}{\small $P_{b_0,l}\cdot \Ltwo^3 / (64\pi L_y\Lone^3)$}}
			\put(-30,25){\small $ T/\Ltwo$}
		\end{subfigure}
	\end{center}
	\caption{\small (\textbf{Top}) Entanglement entropy as a function of the temperature for the theory with $b_0 = 0.6835$ for the different fixed values of the strip width is indicated on each panel. (\textbf{Bottom}) Entanglement pressure as constructed from above case with $ l = 0.14 \Ltwo^{-1}$ with the same value of fixed $b_0$. In all the plots, the dashed vertical line stands for the critical temperature, obtained from thermodynamics. In red lines (dot), the corresponding critical temperature as obtained from the Maxwell construction (entanglement pressure) is indicated. We see that there is good agreement in all cases, within our numerical precision.}\label{fig:FixedWidth_y030}
\end{figure}
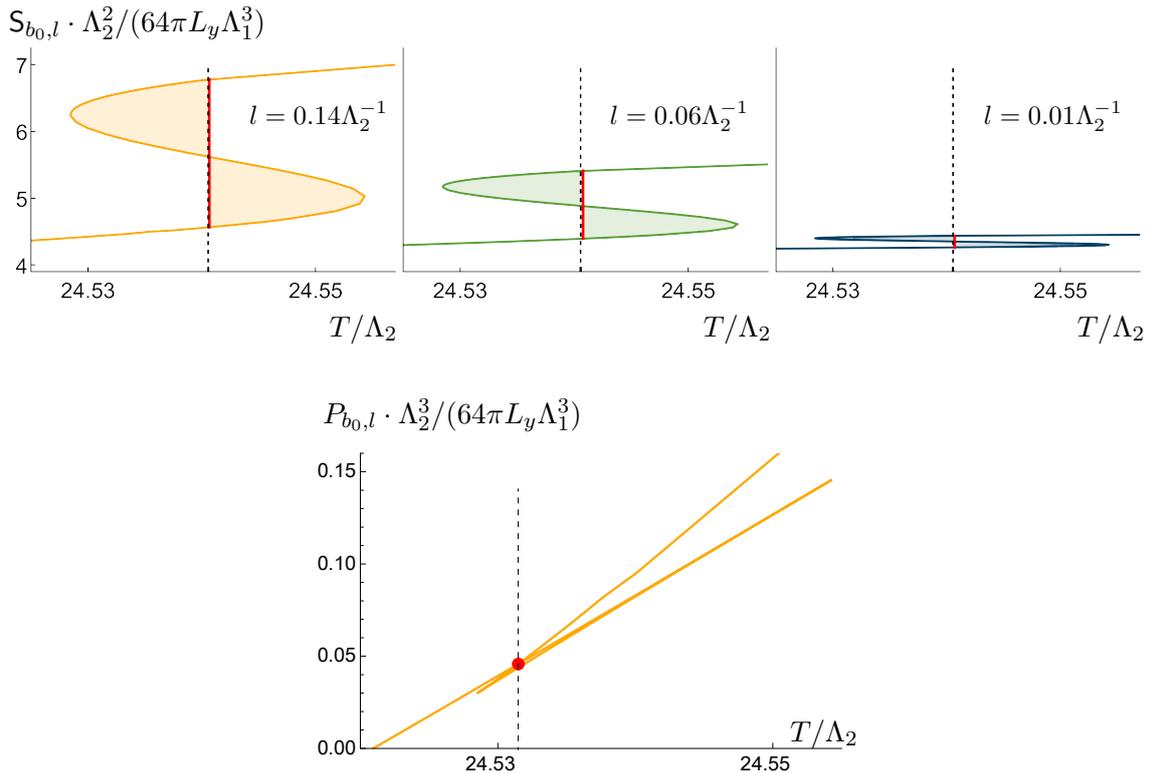

Let us now turn to examine the situation in which the phase transition takes place between two plasma phases, described holographically by two different black brane solutions. This is the case that has deserved more attention in the literature, mainly in bottom-up models \cite{Knaute:2017lll,Asadi:2022mvo}. This case is somehow cleaner than the previous one, for the subtleties of the comparison of entanglement entropy measures between the gapped and the plasma phases being absent.

Indeed, let us choose any reference temperature $T_0$ in Eq.~\eqref{eq:def_pressure}, for which we set $P_{b_0,l}(T_0) = 0$. Even though the actual value of $P_{b_0,l}(T)$ depends on this choice, the difference between the entanglement pressure of two different states does not. In particular, this equation can be used to search for phase transitions: the preferred phase will correspond to the one with the highest pressure.

In Fig.~\ref{fig:FixedWidth_y030} (\textbf{top}) we show $\SEE_{b_0,l}(T)$ for the theory with $b_0\simeq 0.6835$ for different values of the strip width. Some features of this plot are worth mentioning. We notice that, for each choice of the strip width, the corresponding curve for the entanglement entropy displays the ``S'' shape characteristic of a first-order phase transition. Remarkably, this shape is present even for small choices of the strip width. This is interesting because, even though one may have expected that thin strips only probe the UV of the theory, we see that they are still sensitive to the temperature dependence. Geometrically, this is a consequence of the fact that different temperatures correspond to different background geometries. At the same time, we see that the jump in the entanglement entropy approaches zero as the width of the strip vanishes. This assures that we are using the correct counterterms to regularize the entanglement entropy (see Appendix~\ref{ap:strip}) and that the IR (\textit{i.e}, temperature) dependence does indeed fade away in the $l\,\Ltwo\to 0$ limit.

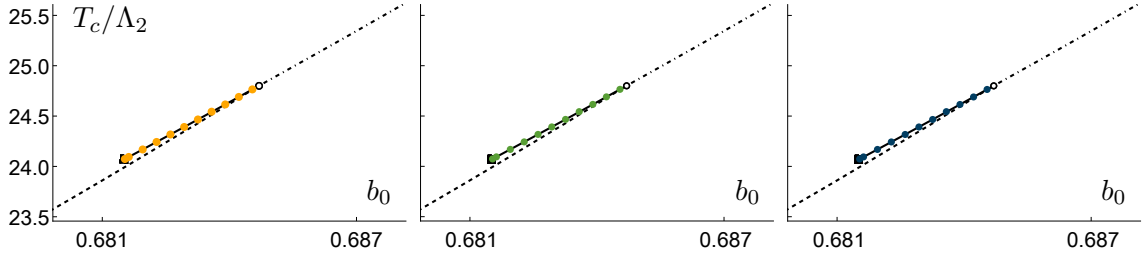
\begin{figure}[t]
	\begin{flushleft}
		\includegraphics[width=\textwidth]{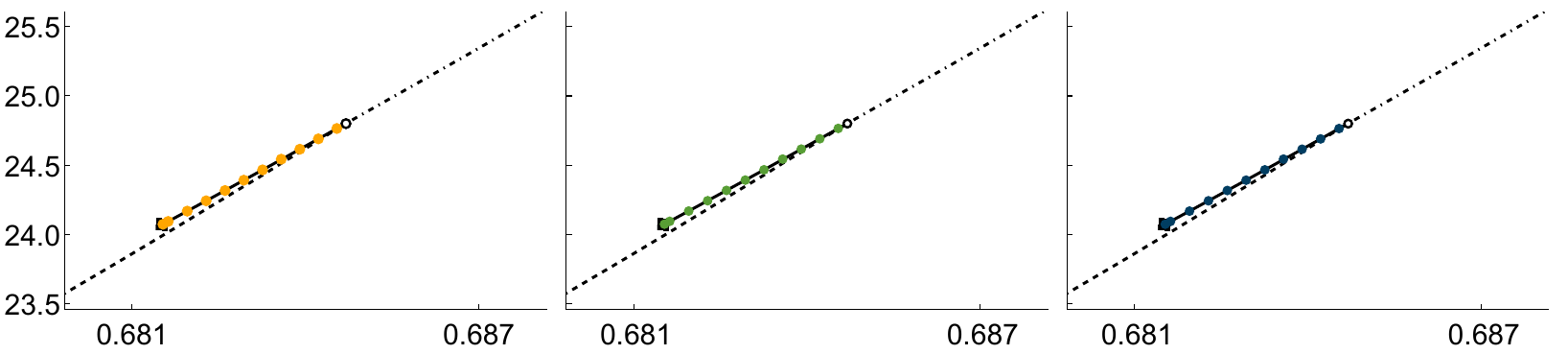} 
		\put(-405,85){\rotatebox{0}{\small$ T_c /\Ltwo $}}
		\put(-295,20){\small$b_0$}
		\put(-157,20){\small$b_0$}
		\put(-20,20){\small$b_0$}
		\caption{\small Position of the phase transition as deduced from the entanglement pressure for different choices of the width, plotted on top of the thermodynamic phase diagram obtained in \cite{Faedo:2017fbv} (see Fig.~\ref{fig:PhaseDiagram}). Here, $ l = 0.06\Ltwo^{-1}$ (left), $0.04\Ltwo^{-1}$ (middle), and $0.005\Ltwo^{-1}$ (right).}\label{fig:widthPT}
	\end{flushleft}
\end{figure}

The properties of the $\SEE_{b_0,l}(T)$ imprints a similarly characteristic swallow-tail shape in the entanglement pressure $P_{b_0,l}(T)$ as computed from Eq.~\eqref{eq:def_pressure}. This can be observed in Fig.~\ref{fig:FixedWidth_y030} (\textbf{down}). From our previous discussion, the arbitrariness in the choice of $P_{b_0,l}(T_0)$ only translates in this plot to the position of the horizontal axes, but it does not alter the values of the temperature where the curves cross or the cusps appear. Thus, we can declare an \textit{entanglement phase transition}\footnote{Recall that this is due to changing external parameters not with varying the subsystem size $l$.} at the temperature for which the two curves cross. We already anticipated in Section~\ref{sec:pressure} that this prescription can be thought of as coming from the Maxwell construction. Indeed, as it can be seen in the figure, the two ways to determine the critical temperature agree to good precision.

Interestingly, not only is this temperature-independent of the choice of the width of the strip, but it also matches, within our numerical precision, the temperature at which the thermal phase transition takes place. This fact can be clearly seen in Fig.~\ref{fig:widthPT}, where the position of these phase transitions is plotted for different theories, and for different choices of the strip width; on top of the thermodynamic phase diagram from Fig.~\ref{fig:PhaseDiagram}. It follows that the line of \textit{type 2} phase transitions is precisely recovered from our entanglement considerations. Note that below $\bcritical$ the curve of $\SEE_{b_0,l}$ ceases to be multivalued and consequently we find no points when $b_0 < \bcritical$. In other words, the transition becomes a crossover.

This contrasts with the results in Ref.~\cite{Knaute:2017lll}. There, the temperature for the phase transition extracted from entanglement considerations was always above the thermal one. We think that there may be two possible reasons for this disagreement. On the one hand, in Ref.~\cite{Knaute:2017lll} the regularization was performed by introducing a UV cutoff in the holographic radial coordinate raising concerns regarding the scheme dependence. For example, it is not clear to us if the cutoff remains temperature-independent. On the other hand, in Ref.~\cite{Knaute:2017lll} the authors did not directly consider the holographic entanglement entropy but a logarithm thereof.

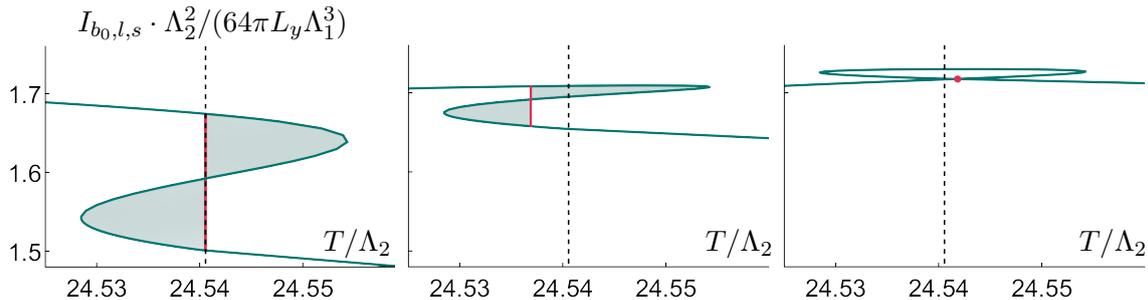
\begin{figure}[t]
	\begin{center}
		\includegraphics[width=\textwidth]{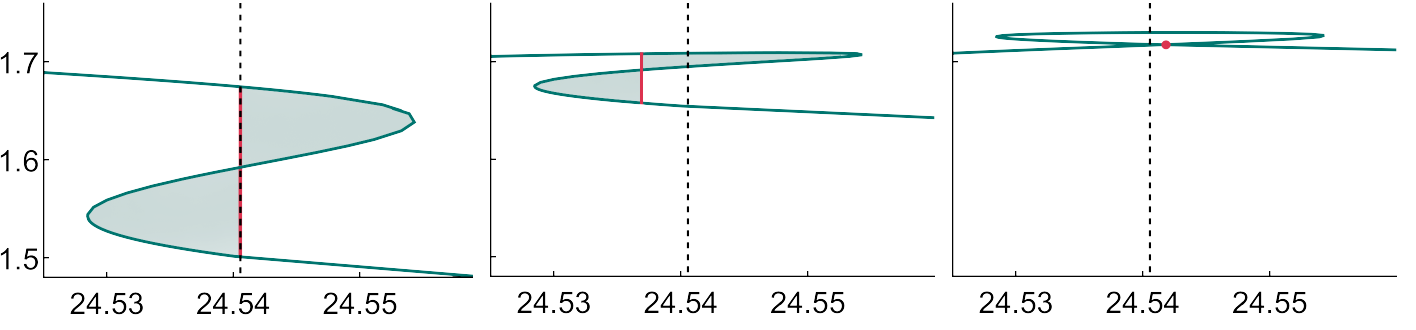} 
		\put(-405,105){\small $I_{b_0,l,s}\cdot \Ltwo^2 / (64\pi L_y\Lone^3)$}
		\put(-313,22){\small $T/\Ltwo$}
		\put(-170,22){\small $T/\Ltwo$}
		\put(-30,22){\small $T/\Ltwo$}
		\caption{\small Mutual information for $b_0 = 0.6835$ with fixed value of $s = 0.005 \Ltwo^{-1}$ and three different fixed values of $l$. (\textbf{Left}) For $l = 0.01 \Ltwo^{-1}$ the mutual information has the familiar shape and can be probed with Maxwell construction. The resulting $T_{EE}$ matches with $T_{thermal} \approx 24.54\Ltwo$. (\textbf{Middle}) For $l = 0.013 \Ltwo^{-1}$ the mutual information curve is already collapsing into a shape where Maxwell construction will not be applicable. (\textbf{Right}) For $l = 0.02 \Ltwo^{-1}$ we encounter the extreme shape of mutual information that is loop-like, and therefore demonstrates how the behavior should not be trusted with larger values of $l$. We conclude that the value $l = 0.01\Ltwo^{-1}$ is the largest at which the results are reliable and the Maxwell construction is applicable; see text for explanation.}\label{fig:MI-looplike}
	\end{center}
\end{figure}

Let us finish making one more remark. The quantity that probed the thermodynamic phase transition in this case has been entanglement entropy, which is not an observable. One could wonder if there is any other quantity that could probe it. For instance, Ref.~\cite{Asadi:2022mvo} studied mutual information and entanglement of purification. Sparked by their works, we comment on the former. Ideally, one would like to define something analogous to the entanglement pressure for the mutual information. Unfortunately, we did not find a working solution for this. Indeed, if we try to implement Maxwell construction with the mutual information, as in Fig.~\ref{fig:MI-looplike}, we see that the critical temperature that is obtained depends on the strip widths and their separations. In particular, it does not match the thermal critical temperature in general.

We can illuminate this by focusing on the formula for the mutual information of strips in Eq.~\eqref{eq:mutualinfostrips}. As prescribed, the mutual information of two strips consists of three different terms. Depending on the chosen values for the strip widths $l$ and the separation $s$ between them, some terms in the formula will dominate over the others. Indeed, for narrow nearby strips the term $S_\cup^{\text{\tiny reg}} (2l+s)$ can be neglected and only the terms proportional to $S_\cup^{\text{\tiny reg}} (l)$ and $S_\cup^{\text{\tiny reg}} (s)$ will be important. In this case, as seen in Fig.~\ref{fig:MI-looplike} (\textbf{left}) Maxwell construction would provide a good approximation of the correct result when trying to locate the phase transition. However, when $l$ is increased, the contribution of the term $S_\cup^{\text{\tiny reg}} (2l+s)$ becomes more and more important. As a consequence the ``S'' shape of the mutual information as a function of the temperature is deformed and eventually disappears, as in Fig.~\ref{fig:MI-looplike} (\textbf{middle}) and (\textbf{right}). In particular, Maxwell construction cannot be used to seek the critical temperature anymore.

In conclusion, for this type of phase transition it seems entanglement pressure is actually providing some insight regarding the thermal phase structure of the system. However, other observables such as mutual information are not as successful in general, even though they can be used in some very specific limits.

\subsection{\textit{Type 3} phase transitions}

\begin{figure}[t]
	\begin{center}
		\begin{subfigure}{.49\textwidth}
			\includegraphics[width=\textwidth]{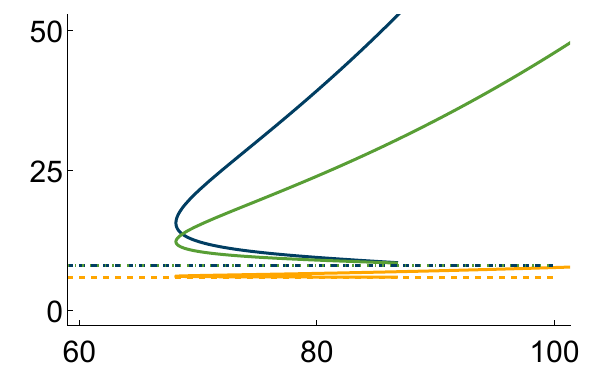} 
			\put(-200,140){\rotatebox{0}{\small $\SEE_{b_0,l}\cdot \Ltwo^2 / (64\pi L_y\Lone^3)$}}
			\put(-25,25){\small $ T /\Ltwo$}
		\end{subfigure}\hfill
		\begin{subfigure}{.49\textwidth}
			\includegraphics[width=\textwidth]{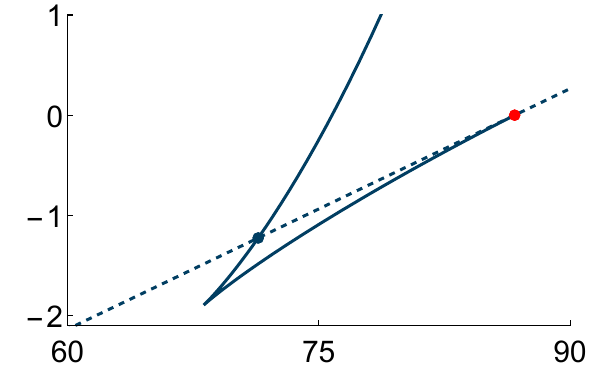}
			\put(-200,140){\rotatebox{0}{\small $10^{-2}P_{b_0,l}\cdot \Ltwo^3 / (64\pi L_y\Lone^3)$}}
			\put(-25,25){\small $ T/\Ltwo$}
		\end{subfigure}
		\caption{\small (\textbf{Left}) Entanglement entropy for fixed $b_0 = 0.6835$ and three different fixed values of $l = 0.04\Ltwo^{-1}$ (dark blue), $0.02\Ltwo^{-1}$ (green) and $0.002\Ltwo^{-1}$ (yellow).  (\textbf{Right}) Entanglement pressure as defined in \eqref{eq:def_pressure} when $l=0.04\Ltwo^{-1}$. The red dot indicates the point where we prescribed that the plasma phase should coincide with the gapped phase.}
		\label{fig:FixedWidth_y0100}
	\end{center}
\end{figure}

Finally, let us discuss how entanglement pressure probes \textit{type 3} phase transitions. In this case, as in Section \ref{sec:type1}, we stumble against the arbitrariness in the definition of entanglement pressure when we want to compare two branches of solutions that are not continuously connected ({\emph{i.e.}}, the gapped and the plasma ones). We will take advantage that whenever a \textit{type 3} phase transition occurs in our system, a \textit{type 1} is also present,\footnote{Note, however, that that for the limiting $b_0\to 1$ case, the \textit{type 1} phase transition is pushed towards $T_c\Ltwo^{-1}\to \infty$.} even though it will be hidden as it is the end of the thermodynamically disfavored branch of black hole solutions, see Fig.~\ref{fig.DifferentCasesThermalPT} (\textbf{bottom}). As in Section~\ref{sec:type1}, we demand that entanglement pressure of both phases has to be the same at the point where the \textit{type 1} phase transition takes place. With this requirement, we can ask how well \textit{type 3} phase transition is probed by this quantity.

For concreteness, let us fix $ l=0.04 \Ltwo^{-1}$ and $b_0 = 0.7902$, and study how the pressure changes with temperature. The result is shown in Fig.~\ref{fig:FixedWidth_y0100} (\textbf{right}). By construction, the curves corresponding to the two phases coincide at the point where the \textit{type 1} phase transition is located. In addition, the two curves cross at a different point, which we take as the probe for the temperature at which the \textit{type 3} phase transition takes place. In this particular case it is $T_c^{\rm EE}( l=0.04 \Ltwo^{-1}) \simeq 71.38 \Ltwo$, which does not coincide with the critical temperature $T_c \simeq 70.94\Ltwo$ Actually, as our notation suggests, $T_c^{\rm EE}(l)$ depends on the width of the strip.

\begin{figure}
    \centering
    \includegraphics[width=.6\textwidth]{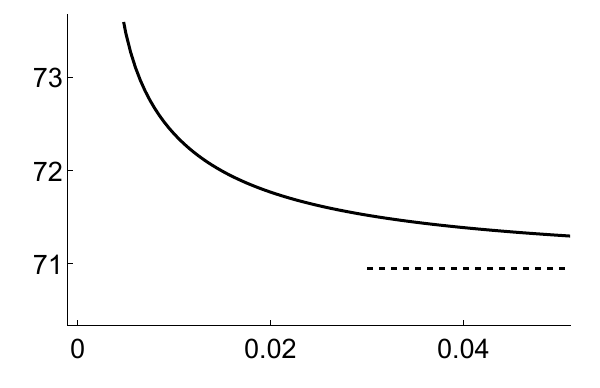}
        \put(10,21){$l \cdot \Ltwo$}
        \put(-290,150){$T_c^{\rm EE}/\Ltwo$}
    \caption{Critical temperature as extracted from entanglement pressure $T_c^{\rm EE}$ for \textit{type 3} phase transitions as a function of the strip width $l$. The dashed line corresponds to the temperature obtained from the thermal phase transitions and appears to be recovered as $l\Ltwo\to\infty$.}
    \label{fig:TcEE-of-width}
\end{figure}

In Fig.~\ref{fig:TcEE-of-width} we show how this quantity changes as a function of the strip width. Moreover, in the limit $l\Ltwo\to\infty$ it appears that the thermal value $T_c$ is recovered.

\subsection{Critical phenomena}

In Section~\ref{sec:type2} we saw that entanglement entropy was quite successful in probing \textit{type 2} phase transitions. In particular, we were able to locate the position of the critical point in the phase diagram. Now, as a last question that we would like to analyze, we ask if it is possible to measure critical exponents of critical phenomena using entanglement quantities.

The critical exponents from the background can be obtained from a formula analogous to Eq.~\eqref{eq:critical_exponentEE}, but replacing the entanglement entropy by the thermal entropy. When applied to our case, our result $\alpha\approx 0.667$ is compatible with the mean-field value $\alpha_\text{mean\ field}={2}/{3}$ stemming from the van der Waals criticality of black holes discussed in Ref.~\cite{Bhattacharya:2017nru}. 
This is not unexpected, because as it happens in field theory \cite{Yaffe:1981vf} the large-$N$ limit suppresses fluctuations.\footnote{Note, however, that there are examples where this is not the case as discussed in Ref.~\cite{Evans:2010np}.}

Let us see if we can get a good estimation of $\alpha$ from our entanglement measures. To do so, we place ourselves near the critical point by choosing a particular value of $b_0\gtrsim \bcritical$ and different strip widths. For this choice, there will still be a first-order phase transition for some $T_c\approx\TCP$. Because of its proximity to the critical point, however, we expect to be able to read off the critical exponents slightly away from~$T_c$.

\begin{figure}[t]
        \begin{center}
		\begin{subfigure}{.49\textwidth}
			\includegraphics[width=\textwidth]{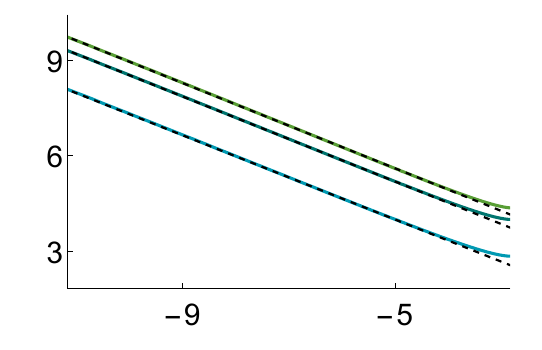} 
			\put(-190,140){\rotatebox{0}{\small $\log\Big[ T \frac{\partial \SEE}{\partial T}\, \cdot \Ltwo^2  (64\pi L_y\Lone^3)^{-1}\Big]$}}
			\put(-85,-5){\small $\log[(T_c-T)/\Ltwo]$}
		\end{subfigure}\hfill
		\begin{subfigure}{.49\textwidth}
			\includegraphics[width=\textwidth]{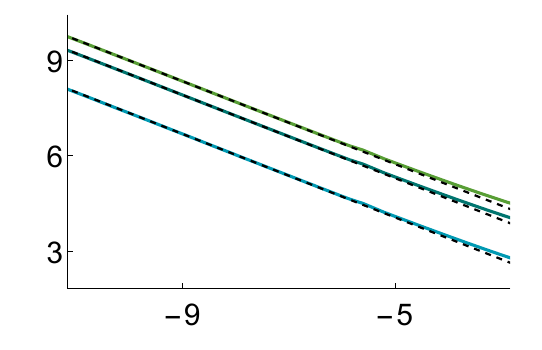}
			\put(-190,140){\rotatebox{0}{\small $\log\Big[ T \frac{\partial \SEE}{\partial T}\, \cdot \Ltwo^2  (64\pi L_y\Lone^3)^{-1}\Big]$}}
			\put(-85,-5){\small $\log[(T-T_c)/\Ltwo]$}
		\end{subfigure}
        \caption{\small Computation of the critical exponents from entanglement entropy. In these plots, we extracted the critical exponent $\alphaEE$ when approaching from the low temperature $T<T_c$ (\textbf{left}) and the critical exponent $\alphaEE'$ when approaching from the high temperature $T>T_c$ (\textbf{right}), using Eq.~\eqref{eq:critical_exponentEE}. We choose a particular value of $b_0\gtrsim \bcritical = 0.6815\ldots$, with three particular choices for the strip width: $l = 0.01\Ltwo^{-1}$ (blue), $0.03\Ltwo^{-1}$  (dark green), and $0.05\Ltwo^{-1}$ (green). The respective exponents $\alphaEE \approx 0.662$ (blue), $0.668$ (dark green), and $0.670$ (green) are obtained from the asymptotic slope of the curves, as indicated by the black dashed straight lines. Similarly, we have $\alphaEE' \approx 0.658$ (blue), $0.657$ (dark green), and $0.652$ (green).}
    \label{fig:CriticalExponents}
    \end{center}
\end{figure}

Indeed, that is the case, as can be seen in Fig.~\ref{fig:CriticalExponents}. There, a value of $b_0$ which is very close to $\bcritical$ is chosen, and the analysis is performed for three different choices of the strip width. As we can see in the figure, the value of the exponent in each case is indeed $\alphaEE \approx 2/3$, which matches the thermal value and agrees with the expected critical exponents.

Let us finish this section with one remark. Recall that the mutual information did not probe the phase transition correctly in general (see Fig.~\ref{eq:mutualinfo}). However, we observed in Section~\ref{sec:type2} that it approximately did so in the limit of narrow strips. A consequence of this is that critical exponents can also be read off from an equation analogous to Eq.~\eqref{eq:critical_exponentEE}, with entanglement entropy $\SEE_{b_0,l}(T)$ substituted by mutual information $I(b_0,T,l,s)$ between narrow strips.\footnote{In Ref.~\cite{Asadi:2022mvo} the critical exponents were computed both from the holographic mutual information of strips and the holographic entanglement of purification.} We have explicitly checked that the holographic mutual information leads to the same critical exponents.

\section{Discussion}\label{sec:discussion}

In this work we demonstrated that the entanglement entropy will provide a complementary tool to study characteristics of phase transitions in strongly coupled gauge theories. We focused on a family of theories endowed with interesting IR properties and a rich phase diagram uncovered in Ref.~\cite{Elander:2020rgv}. In particular, at finite temperature, we find three types of phase transitions of different nature, that meet at a triple point. The phase diagram is also endowed with a critical point where a second-order phase transition occurs, at the end of a line of first-order phase transitions.

Considering the entanglement entropy of strips at different temperatures, we defined \textit{entanglement pressure}, mimicking the thermal case. The crossing of this pressure with itself was able to probe phase transition between plasma phases, dual to black brane solutions on the gravity side. However, it turned out less successful in detecting phase transitions between plasma and gapped phases. This is one of the main messages of this work: The description of the transition between deconfined and gapped (or confining) phases using entanglement measures remains unsatisfactory. We invite experts in addressing this conundrum in any setup, especially when there is a dual gravity description.

On the contrary, we discovered that entanglement entropy can be successfully used to locate the critical point. In particular, the corresponding entanglement specific heat accurately predicts the corresponding critical exponent. We believe that this result is quite general and should be scrutinized beyond strongly coupled (holographic) gauge field theories. To this end, recall that the entanglement specific heat involves derivatives, meaning that it is scheme-independent, and that it is to be analyzed with fixed subsystem sizes, thereby amenable to real world studies. Support to our claim comes from lattice Yang-Mills theory, where the correct value for the critical exponent as extracted from entanglement was indeed recovered \cite{Jokela:2023yun}.

Finally, because of scheme-dependence of the entanglement entropy, we resorted to also other observables. We attempted to extend the Maxwell construction to mutual information for non-overlapping strips to analyze the properties of phase transitions. We found that mutual information fails as a probe of phase transitions in general, except for a narrow range in the parameter space.

To find an entanglement measure that probes phases of matter faultlessly in general, {\emph{i.e.}}, at least for all types of phase transitions unraveled in this work, is a very interesting open problem.

\begin{acknowledgments}
We thank Ant\'on Faedo for valuable discussions. J.~S. thanks the possibility for participating in the PiTP 2023 program: \textit{``Understanding Confinement''} in the last stages of this project. There, he enjoyed discussions on the topic of entanglement entropy as a probe of confinement with Andrea Bulgarelli, Igor Klebanov, and Pedro Jorge Martinez, to whom he is truly thankful. Nordita is supported in part by NordForsk. H.~R. is supported in part by the Finnish Cultural Foundation.
\end{acknowledgments}

\appendix

\section{Details of the background solutions}\label{ap:solution}

In this appendix we gather the relevant technical details of the system studied in this paper and the background solutions considered. For a detailed explanation, see Refs.~\cite{Faedo:2017fbv,Elander:2020rgv}.

The internal compact space of the solutions consists of a squashed $\CP^3$ and the geometry asymptotes in the UV to that of a stack of $N$ D2-branes. For this reason it is convenient to consider the Ansatz
\be\begin{aligned}
    \dd s^2_{\text{\tiny st}} &= h^{-\frac{1}{2}}\left(-\mathsf{b} \dd t^2 + \dd x_1^2+\dd x_2^2\right)+h^{1/2}\left(\frac{\dd r^2}{\mathsf{b}}+e^{2f}\dd \Omega_4^2 +e^{2g}\left[\left( E^1\right)^2 +\left(E^2\right)^2\right] \right)\\
    e^{\Phi} &= h^{\frac{1}{4}}e^\Lambda
\end{aligned}
\ee 
for the metric and the dilaton field. The dilaton and all the metric functions $f,\, g,\, \Lambda,\, h$ depend only on the radial coordinate $r$. Moreover, the complex projective plane is considered to be partioned as the quotient space Sp($2$)$/$U($2$), consisting of a two-sphere
(described by the vielbeins $E^1$ and $E^2$) fibered over a four-sphere with metric $\dd \Omega_4^ 2$. We parametrize $\CP^3$ as in Refs.~\cite{Conde:2011sw,Jokela:2012dw}. Considering a set of left-invariant one-forms on the three-sphere $\omega^1$, $\omega^2$, and $\omega^2$, the metric of the four-sphere with unit radius can be written as
\be 
    \dd\Omega_4^2 = \frac{4}{(1+\xi^2)^2}\left(\dd\xi^2 + \frac{\xi^2}{4}\omega^i\omega^i \right)
\ee 
with $\xi\in(0,\infty)$ a non-compact coordinate. Now we can introduce two angles $\theta\in(0,\pi)$ and $\varphi\in(0,2\pi)$ to parametrize the two-sphere. The non-trivial fibration appears manifest in the expression of the vielbens
\begin{align}
    E^1 &= \dd\theta + \frac{\xi^2}{1+\xi^2}(\sin\varphi\omega^1-\cos\varphi\omega^2)\\
    E^2 &= \sin\theta\left(\dd\varphi -\frac{\xi^2}{1+\xi^2}\omega^3 \right)+\frac{\xi^2}{1+\xi^2}\cos\theta(\cos\varphi\omega^1 + \sin\varphi\omega^2) \ .
\end{align}
It is convenient to consider the following rotated version of the vielbeins on the four-sphere 
\begin{align}
    \mathcal{S}^1 &= \frac{\xi}{1+\xi^2}[\sin\varphi\omega^1 -\cos\varphi\omega^2]\\
    \mathcal{S}^2 &= \frac{\xi}{1+\xi^2}[\sin\theta\omega^3 -\cos\theta(\cos\varphi\omega^1 + \sin\varphi\omega^2)]\\
    \mathcal{S}^3 &= \frac{\xi}{1+\xi^2}[\cos\theta\omega^3 +\sin\theta(\cos\varphi\omega^1 +\sin\varphi\omega^2)]\\
    \mathcal{S}^4 &= \frac{2}{1+\xi^2}\dd\xi \ .
\end{align}
Despite the explicit dependence on the angles $\theta$ and $\varphi$, it holds that $\mathcal{S}^n\mathcal{S}^n= \dd\Omega_4^2$. Now, it is possible to write down the left-invariant forms on the quotient space in terms of these vielbeins. These contain the two-forms
\begin{equation}
    X_2 = E^1\wedge E^2\ ,\qquad J_2 = \mathcal{S}^1\wedge \mathcal{S}^2 = \mathcal{S}^3\wedge \mathcal{S}^4
\end{equation}
and the three-forms
\begin{eqnarray}
X_3&=&E^1\wedge\left(\mathcal{S}^1\wedge\mathcal{S}^3-\mathcal{S}^2\wedge\mathcal{S}^4\right)-E^2\wedge\left(\mathcal{S}^1\wedge\mathcal{S}^4+\mathcal{S}^2\wedge\mathcal{S}^3\right)
\nonumber\\[2mm]
J_3&=&-E^1\wedge\left(\mathcal{S}^1\wedge\mathcal{S}^4+\mathcal{S}^2\wedge\mathcal{S}^3\right)-E^2\wedge\left(\mathcal{S}^1\wedge\mathcal{S}^3-\mathcal{S}^2\wedge\mathcal{S}^4\right)\,;
\end{eqnarray}
related by exterior differentiation
\begin{equation}\label{eq:relationForms}
\dd X_2\,=\,\dd J_2\,=\,X_3\ ,\qquad \dd J_3\,=\,2\left(X_2\wedge J_2+J_2\wedge J_2\right)\ .
\end{equation}
Higher-rank left-invariant forms can be constructed by wedging these. In particular, we find two four-forms $X_2\wedge J_2$ and $J_2\wedge J_2$ and the volume form of $\CP^3$, $\Omega_6 =  (E_1 \wedge E_2)\wedge (\mathcal{S}^1\wedge\mathcal{S}^2\wedge\mathcal{S}^3\wedge\mathcal{S}^4)$. There are no adequate one- or five-forms and the complete set closes under Hodge duality.

Our conventions for type IIA supergravity are such that the Bianchi identities for the forms read
\begin{equation}\label{eq:IIABianchis}
 \dd H_3=0\ ,\,\qquad\qquad \dd F_2=0\ ,\qquad\qquad \dd F_4=H_3\wedge F_2\,,
\end{equation}
while the string-frame equations of motion are
\begin{equation}\label{eq:IIAeoms}
\begin{aligned}
\dd*F_4+H_3\wedge F_4 &=  0  \\ 
\dd*F_2+H_3\wedge * F_4 &= 0 \\
\dd\left(e^{-2\Phi}*H_3\right) -F_2\wedge*F_4-\frac12 F_4\wedge F_4&= 0\ .
\end{aligned}
\end{equation}

A convenient Ansatz for the forms is
\begin{equation}\label{fluxesansatz}
\begin{array}{rclcrcl}
F_4 &=&  \mathbf{f}_4 *\Omega_6 + G_4 + B_2\wedge F_2 \,,&\qquad& F_2& =& Q_k (X_2 - J_2)\,,\\[2mm]
 G_4 &=& \dd(a_J J_3) + q_c\left(J_2\wedge J_2 - X_2\wedge J_2\right)\,,&\qquad & B_2& =& b_X X_2 + b_J J_2\,,
 \end{array}
\end{equation}
where we have defined the quantity\begin{footnote}{ \label{footnote:Qc}
Note that the constant $Q_c$ is related to the number of $N$ branes through the standard quantization condition \cite{Faedo:2017fbv,Elander:2020rgv}. However, this statement needs closer inspection. In Ref.~\cite{Faedo:2022lxd}, it was noted that for the confining case ($b_0=1$) $Q_c$ must be taken to be zero, as it accompanies a collapsing circle. Then, one should identify the number of branes $N$ from the asymptotic growth of the warp factor $h$ instead, see Eq.~\eqref{eq:toD2intheUV}. Considering solutions with $Q_c\neq0$ still makes sense as they are related through a large gauge transformation, which implements a duality cascade. The corresponding analysis for the non-confining case ($Q_k\neq0$, $b_0<1$), see Ref.~\cite{Hashimoto:2010bq}, needs to be performed with the eleven-dimensional uplifted solution and has not yet been worked out in our conventions.}
\end{footnote}
\begin{equation}
 \mathbf{f}_4= Q_k b_J^2 +2(q_c+2a_J)b_X - 2 b_J(q_c-2a_J+Q_kb_X)+Q_c\,.
\end{equation}
The parameters $Q_c$, $q_c$, and $Q_k$ are constants related to gauge theory parameters
\begin{equation}\label{gaugeparam}
Q_c\,=\,3\pi^2\ell_s^5 g_s\,N\,,\qquad\qquad 
q_c\,=\,\frac{3\pi\ell_s^3g_s}{4}\,\parent{M-\frac{k}{2}}\,,\qquad\qquad Q_k\,=\,\frac{\ell_sg_s}{2}\,k\,.
\end{equation}
These parameters were grouped together with the gauge coupling $\lambda = \ls^ {-1} \gs N$ to construct the scales $\Lone$ and $\Ltwo$ in Eq.~\eqref{eq:units}. One should note that $N$, $M$, and $k$ are not completely independent of the choice of $b_0$, as pointed out in Ref.~\cite{Hashimoto:2010bq}. As a consequence, one expects that $\Lambda_1$ and $\Lambda_2$ will be related to $b_0$, and consequently to the different gauge couplings $g_1$, $g_2$ in Eq.~\eqref{eq:b0.coupligs}. This analysis, that requires understanding of how the cascade is realized in eleven-dimensonal supergravity (see footnote \ref{footnote:Qc}), will be worked out somewhere else.

On the other hand, $b_J$, $b_X$, and $a_J$ depend on the radial coordinate, their dynamics dictated by the form Eqs.~\eqref{eq:IIAeoms}. Together with these, we will also have to solve the equations of motion for the dilaton
\begin{eqnarray}
R+4\nabla_M\nabla^M \Phi - 4 \nabla^M\Phi \nabla_M \Phi-\frac{1}{12} H^2 = 0\,,
\end{eqnarray}
and the metric
\begin{equation}
R_{MN} + 2 \nabla_M\nabla_N \Phi -\frac{1}{4} H_{MN}^2 = e^{2\Phi}\left[
\frac{1}{2} (F_2^2)_{MN} + \frac{1}{12}(F_4^2)_{MN} - \frac{1}{4}g_{MN}\left(
\frac{1}{2}F_2^2 +\frac{1}{24}F_4^2\right)\right]\,.
\end{equation}

\section{Computation of the entanglement entropy of the strip}\label{ap:strip}

In this appendix, we go through details on the computation of the entanglement entropy of the strip and discuss how counterterms are properly taken into consideration. We recover here the expression for the entanglement entropy in Eq.~\eqref{eq:EEconnectedConf} for ease of reference
\begin{equation}\label{eq:EEconnectedConf_in_Appendix}
    \SEE_\cup(b_0,T,l) = \frac{V_6 L_y}{4 G_{10}}\int_{-\frac{l}{2}}^ {\frac{l}{2}}\dd \sigma^1 \,   \Xi^{\frac{1}{2}} \left[1+\frac{h}{\mathsf{b}}\dot{r}^2\right]^{\frac{1}{2}}\,.
\end{equation}
Recall that $\Xi(r) = h^2e^{8f+4g-4\phi}$. As we mentioned, this integral enjoys a conserved quantity that permits to find a simple expression for the embedding,
\begin{equation}
    \label{eq:embedding}
    \dot{r} = \pm \sqrt{\frac{\mathsf{b}}{h}} \, \sqrt{\frac{\Xi\ }{\Xi_*} - 1} \ ,
\end{equation}
where $\Xi = \Xi(r_*)$ and the dot stands for differentiation with respect to $\sigma_1$. With this we can rewrite the entanglement entropy of the strip as a simple integration of metric functions
\begin{equation}
    \SEE_\cup(b_0,T,l) = 2 \frac{V_6 L_y}{4 G_{10}} \int_{r_*}^{\Lambda|Q_k|} \dd r \,  \frac{\Xi }{\sqrt{\Xi-\Xi_*}} \sqrt{\frac{h}{\mathsf{b}}}\ .
\end{equation}
However, as we mentioned in Section~\ref{sec:EE}, this quantity is UV divergent and needs to be renormalized. That is why we have explicitly introduced the UV cut-off $\Lambda$, which will eventually be taken to infinity after regularization.

The UV expansions of the metric functions and the dilaton where worked out in Ref.~\cite{Elander:2020rgv}. Following them, it will be easy to work out the divergence structure of the entanglement entropy ${S}_\cup$. Let us first write this quantity in the $u$ coordinate defined below Eq.~\eqref{eq:UVexp}
\begin{equation}\label{eq:UVofEE1}
    \SEE_\cup(b_0,T,l)=  2 |Q_k| \frac{V_6 L_y}{4 G_{10}} \int_{u_*}^{\Lambda^{-1}} \dd u \, \left( -\frac{1}{u^ 2} \frac{\Xi }{\sqrt{\Xi-\Xi_*}} \sqrt{\frac{{h}}{\mathsf{b}}}\ \right) \equiv \int_{u_*}^{\Lambda^{-1}} \dd u \,  L_\cup\,,
\end{equation}
where $u_* = |Q_k| r_*^ {-1}$ is the value of the radial variable $u$ at the turning point. The last equality in Eq.~\eqref{eq:UVofEE1} defines $L_\cup$. Replacing the functions in terms of their UV expansion and performing the indefinite integral we obtain 
\begin{equation}\label{eq:UVofEE}
    \begin{aligned}
    \SEE_\cup &=  2 \,\frac{64\pi L_y\Lone^3}{\Ltwo^2} \Bigg[ \frac{1}{30} \left(1-b_0^2\right) \Lambda ^2-\frac{4}{45} \left(4 b_0^2+1\right) \Lambda - \frac{4}{63} \left(21
   b_0^2-1\right) \log \Lambda + S_0 \\&\qquad +
   \frac{64}{315} \left(21 b_0^2-1\right) \Lambda^{-1} +
   \frac{64}{189} \left(21 b_0^2-1\right) \Lambda^{-2} + 
   \OO(\Lambda^{-3})\Bigg]\,.
    \end{aligned}
\end{equation}

Let us pause and comment a little bit on this result. First, note that the leading order divergence in Eq.~\eqref{eq:UVofEE} appears at the same order in the cutoff as in Ref.~\cite{vanNiekerk:2011yi}. Secondly, it is easy to read off the necessary counterterms to render the entanglement entropy finite from this last expression,
\begin{equation}\label{eq:Sct}
    \SEE_{\text{\tiny ct}} (b_0,\Lambda^{-1}) =   2 \,\frac{64\pi L_y\Lone^3}{\Ltwo^2} 
 \Bigg[\frac{1}{30} \left(1-b_0^2\right) \Lambda ^2-\frac{4}{45} \left(4 b_0^2+1\right) \Lambda - \frac{4}{63} \left(21
   b_0^2-1\right) \log \Lambda\,\Bigg].
\end{equation}
Interestingly, the expression is proportional to the counterterms used in Ref.~\cite{Jokela:2020wgs} for the entanglement entropy of a disk (up to a factor of the radius of the disk). Note that it does not depend on the temperature $T$, nor the strip width. Rather, as it should, it only depends on UV data. Related to this, note that any constant could be added to Eq.~\eqref{eq:Sct} that would combine with $S_0$ in Eq.~\eqref{eq:UVofEE}. This constant, however, would spoil the comparison between different states if it were, for example, temperature-dependent. Actually, this dependence could unintentionally be introduced through a temperature-dependent coordinate $\tilde u$, inducing a  temperature dependence in the corresponding cutoff $\tilde\Lambda$. Using the usual relation $U = r/\ell_s^2$ \cite{Itzhaki:1998dd} to translate the original radial coordinate to a gauge theory energy scale $U$, we see that $\Lambda$ is related to a temperature-independent UV energy scale measured in units of the gauge coupling
\begin{equation}
    \Lambda = \frac{r_{\text{\tiny UV}}}{|Q_k|} = \frac{2N \, U_{\text{\tiny UV}}}{\lambda |k|}\,.
\end{equation}
Thus, Eq.~\eqref{eq:Sct} is temperature-independent. 

Considering that the counterterms can be written as
\begin{equation}
\label{eq:regFromLagrangian}
\begin{aligned}
\SEE_{\text{\tiny ct}} (b_0,\Lambda^{-1}) = \int_{u_*}^{\Lambda^{-1}}\dd u \ &L_\cup^{\text{ct}}  \ + \SEE_{\text{ct}} (b_0,u_*)\\
\mbox{with }\qquad &L_\cup^{\text{ct}}  \equiv  2 \,\frac{64\pi L_y\Lone^3}{\Ltwo^2} \left[\frac{b_0^2-1 }{15 u^3}+\frac{4 \left(4 b_0^2+1\right) }{45
	u^2}+\frac{4 \left(21 b_0^2-1\right) }{63 u}\right]\ ,\\[2mm]
\end{aligned}\end{equation}
the expression for the regularized entanglement entropy of the strip can be rearranged in a way that facilitates numerical computations, namely
\begin{equation}
    \SEE_\cup^{\text{\tiny reg}} = 
    \lim_{\Lambda\to\infty}\left[ \SEE_\cup - \SEE_{\text{\tiny ct}}(b_0,\Lambda^{-1}) \right] =
    \int_{u_*}^0 \dd u \, \left( L_\cup - L_\cup^{\text{ct}}  \right) - \SEE_{\text{\tiny ct}} (b_0,u_*)\,.
\end{equation}


\bibliographystyle{JHEP}
\bibliography{references}
\end{document}